\documentclass[12pt,preprint]{aastex}
\slugcomment{Accepted  for publication in 
 {\it Astronomy and Astrophysics} on July 8, 2017}  
\def\lax {\ifmmode{_<\atop^{\sim}}\else{${_<\atop^{\sim}}$}\fi}  
\def\gax {\ifmmode{_>\atop^{\sim}}\else{${_>\atop^{\sim}}$}\fi}  
\def\gtorder{\mathrel{\raise.3ex\hbox{$>$}\mkern-14mu
             \lower0.6ex\hbox{$\sim$}}}

\def\cm2{cm$^{-2}$}
\def\s1{s$^{-1}$}

\begin{document}

%\title{4U~1705-44:A Hybrid  {\it atoll} and {\it Z} Type X-ray Binary
% as a %(link between) 
%unique hybride type of atoll and {\it Z} source properties
%}
%\title{Stability of the photon indices in atoll-source 4U~1705-44 for spectral states 
%transitions
%}

\title{%{\it Beppo}SAX, {\it Suzaku} and {\it RXTE} 
Swift J164449.3+573451 and Swift J2058.4+0516: Black hole mass estimates for  tidal disruption event sources
%. SPECTRAL HARDENING DURING the BANANA BRANCH
}

% Observational Evidence for Neutron Star in GX~340+0

%\title{The stability of spectral index of the ``hard component'' as a function of mass accretion rate in Z-source GX~340+0}
% Observational Evidence for Neutron Star in GX~340+0

%\title{On the Constancy of the Photon Index of  X-ray spectra of 4U~1728-34 through all spectral states} 
%during outburst transitions}

\author{ Elena Seifina\altaffilmark{1,2}, Lev Titarchuk\altaffilmark{3,4},   
\& Enrico Virgilli\altaffilmark{3}}
%\& Nikolai Shaposhnikov\altaffilmark{3} }
%\author{Elena Seifina\altaffilmark{1}, Lev Titarchuk\altaffilmark{2}}%  \& Filippo Frontera\altaffilmark{3} }
%\altaffiltext{1}{Moscow State University/Sternberg Astronomical Institute, Universitetsky 
%Prospect 13, Moscow, 119992, Russia; seif@sai.msu.ru}
\altaffiltext{1}{LAPTh, Annecy-le-Vieux F-74941, France; seifina@lapth.cnrs.fr}
\altaffiltext{2}{Moscow State University/Sternberg Astronomical Institute, Universitetsky 
Prospect 13, Moscow, 119992, Russia; seif@sai.msu.ru}
\altaffiltext{3}{Dipartimento di Fisica, Universit\`a di Ferrara, Via Saragat 1, I-44122 Ferrara, Italy, email:titarchuk@fe.infn.it, email; email:virgilli@fe.infn.it}
\altaffiltext{4}{National Research Nuclear University MEPhI (Moscow Engineering Physics Institute), Moscow, Russia}

%ICRANET, Piazza della Repubblica 10-12 65122 Pescara,  Italy; 
%  Goddard Space Flight Center, NASA,  code 663, Greenbelt  
%MD 20770, USA; email:lev@milkyway.gsfc.nasa.gov, USA}
%\altaffiltext{3}{NASA Goddard Space Flight Center, NASA, Astrophysics Science Division, Code 661, Greenbelt, MD 20771, USA; Chris.R.Shrader@nasa.gov}
%\altaffiltext{4}{Universities Space Research Association, 10211 Wincopin Cir, Suite 500, Columbia, MD 21044, USA}
%\altaffiltext{3}{CRESST/University of Maryland, Department of Astronomy, College Park, MD 20742, USA} 
%Goddard Space Flight Center, NASA,  code 663, Greenbelt  
%MD 20771,  USA: email:nikolai.v.shaposhnikov@nasa.gov}
%\altaffiltext{3}{Dipartimento di Fisica, Universit\`a di Ferrara, Via Saragat 1, I-44122  Ferrara, Italy, email:frontera@fe.infn.it
%}

\begin{abstract}
{A tidal disruption event (TDE) is an astronomical phenomenon in which { a} previously dormant black hole 
(BH)  destroys a star passing too close to its central part. We analyzed
%spectral analysis of
 the flaring episode %phase X-ray data 
detected from  the TDE sources,   %bright point source flare in 
Swift J164449.3+573451  and Swift J2058.4+0516 (hereafter Swift~J1644+57 and Swift J2058+05, respectively) 
using {\it RXTE}, {\it Swift} and  {\it Suzaku} data.  
%for each of the sources. { For Swift~J1644+57  source we also used $Suzaku$ data (2011, 2012).} 
The spectra are well fitted by the so called {Bulk Motion Comptonization} model for which  the best-fit photon index $\Gamma$ varies from {1.1} %1.1 
to 1.8. We have firmly established the saturation of the photon index versus mass accretion rate %, $\dot M$ 
at  $\Gamma_{sat}$  about 1.7 -- 1.8.  
The saturation  of  $\Gamma$  is usually identified as a signature of a BH now established  in Swift~J1644+57 and Swift J2058+05. 
In Swift~J1644+57 we found the relatively low $\Gamma_{sat}$ values which indicate { a} %to  the 
high electron (plasma) temperature,  $kT_e\sim$ 30 -- 40 keV. {This} %,  which  
is  also consistent with  high cutoff energies, $E_{cut}\sim$ 60 -- 80 keV found  using best fits of   the $RXTE$  spectra.  Swift~J2058+05  shows a lower electron  
temperature, $kT_e\sim$ 4$-$10 keV than that for Swift~J1644+57. 
For the BH mass estimate we used the scaling technique taking  the Galactic BHs, GRO J1655--40, 
{ 
GX~339--4, Cyg~X--1 and 4U~1543--47
}
as reference sources and found that the BH mass in 
{Swift~J1644+57} %this source 
is $M_{BH}\ge 7\times 10^6$ M$_{\odot}$ assuming the distance to {this source} %Swift~J1644+57 %Swift J1644449.3+573451 
of 1.5 Gpc. 
%For the BH mass estimate in Swift~2058+05 we use the same scaling method  taking  the Galactic BH, GRO J1655--40 and extragalactic BH source Swift~1644+57 as reference sources and found that the BH mass in 
For {Swift~J2058+05} %, 
{we obtain}  $M_{BH}\ge 2\times 10^7$ M$_{\odot}$ assuming the distance to {this source} %Swift~J1644+57 %Swift J1644449.3+573451 
of 3.7 Gpc.
We have also found that the seed (disk) photon temperatures are quite low, of order of 
100 -- 400 eV, in both of the sources,  which are  consistent with the  estimated BH masses. } 
\end{abstract}


\section{Introduction}
Rees (1988), hereafter R88,  suggested 
%As was suggested by Martin Rees (1988) 
that among bright X-ray flares that continue  for a few years,  
some can be caused by  tidal  gravitational disruption
of a star which  passed  too close  to a super massive black hole (SMBH). If a star passes
within the tidal radius of a black hole
%less than 100 Schwarzschild radii of a SMBH,
then the  gravity   rips the star apart (R88). As the stellar remnant
approaches a black hole (BH), its gravitational potential energy is converted into heat
through viscous effects. Some of the debris can be ejected, while the remaining part can be 
ingested by a central SMBH. In this case the accretion flow  reaches a temperature of about $10^5$ K and  emits brightly at optical, ultraviolet, and X-ray wavelengths during the period from  about 100 days to a few years. 
% TDEs are phenomenon associated with the 
%or tidal disruption of stars by a nearby black hole, hosting e.g. in the center of galactic nucleus. 
%The tidal disruption event (TDE), as suggested,  occurs  at the time scale  related to  quasar or stellar mass  evolution (R88).  
\cite{Komossa15} described  observational appearances of these events in detail. 

Swift~164449.3+573451  (hereafter Swift  J1644+57)  was initially  discovered   as a $\gamma$-ray burst (GRB) event.
The $Swift$ Burst Alert Telescope (BAT, 15 -- 150 keV) detected  a new
uncataloged source on 2011 March 28. % in a 1208 s Image Trigger beginning at T0 = 12:57:45 UT33. 
Because the source was assumed to be a GRB, it was named as GRB 110328A. %, following standard nomenclature. 
However, this first detection was also followed by three additional flares with a subsequent 
increase of {\bf the} count rate over the next two days and thus, it  was   recognized that this object was not a GRB. 
%As a result, 
Now, the name of the source, Swift~J1644+57 
%was renamed as Swift J164449.3+573451  (Swift~J1644+57), 
%which is  
based on the initial position of its X-ray counterpart { is commonly used}. 

Swift~J1644+57 is also characterized by the long duration of the X-ray outburst %emission 
(with a power-law decay $\propto t^{-5/2}$) and specific flaring events. % are  characteristics of Swift~J1644+57. 
We also  note  that
the spatial coincidence of it  with the central nucleus of a galaxy at redshift 0.354 (the luminosity
distance, 1.5 Mpc, see %.5.7 _~ 1027 cm; 
Levan et al. 2011) indicates  that this source was not a GRB, but associated 
with an  accretion event  in the close vicinity  of
a  SMBH (see also Bloom et al. 2011; Burrows et al. 2011).

We should emphasize that this  TDE phenomenon is only occasionally  observed and therefore is poorly  studied. Furthermore, there are no direct  measurements of the mass, $M_{BH}$ of a dormant BH except of our Galactic center black hole.  In the Swift~J1644+57 case  the optical emission lines imply that the host object is not 
an active galactic nucleus (AGN)
%\LEt{ Please introduce the acronym at first use, both in the abstract and again in the main text.}, 
but a HII-type galaxy. Variable emission was also detected in the near-IR and was not detected in the optical band (possibly due to the excess of extinction in the 
optical spectrum). Various estimates of a BH mass, based on  the %most 
rapid variability 
timescale, all give that  $10^7 < M_{BH} < 10^9$ M$_{\odot}$. % $M_{BH}<10^7$ M$_{\odot}$. 
This indirect %order-of-magnitude 
constraints provide only  %consist of 
an upper limit to the BH mass  (see R88), 
for example, a minimum variability time %-scale 
of $\sim$ 100 s that sets an upper limit to the light-crossing time %-scale 
(Bloom et al. 2011; Burrows et al. 2011; {Liang \& Liu 2003}) and well-known empirical relations between $M_{BH}$ and the 
host galaxy environment %(whatever causes those trends; e.g. 
(Silk \& Rees 1998; Jahnke \& Maccio 2011).  Based on  optical 
luminosity of the host galaxy, Levan et al. (2011) evaluated a spheroidal  mass, $M_{sph}\sim 10^9 - 10^{10}$ 
$M_{\odot}$. Using the log-linear spheroidal mass and the BH mass relation
of Bennert et al. (2011), %(see also Magorrian et al. 1998; Lauer et al. 2007; Kormendy, Bender \& Cornell 2011), 
Levan et al.  found that  likely a BH mass within  2$\times 10^6 \le M_{BH} \le 10^7$ $M_{\odot}$ taking into account the  relation by Graham (2012) 
instead  of the relation, $M_{sph}-M_{BH}$.  The latter relation %It 
also provides a lower range of BH masses, 
$10^5 M_{\odot}\le M_{BH} \le 10^7$ $M_{\odot}$ which  is more consistent with the BH mass, 
$\log(M_{BH}/M_{\odot}) = 5.5\pm 1.1$ estimated by Miller \& Gultekin (2011), based on empirical, so called  fundamental plane relations between radio and X-ray luminosities of accreting BHs.
Many subsequent observations of Swift~J1644+57 have been performed at all wavelengths 
(X-rays: Mangano et al., 2016; Gonzalez-Rodriguez, 2014; Castro-Tirado  et  al., 2013; Zauderer  et  al.,  2013;  Saxton et al., 2012; Reis et  al.,  2012; radio: Cendes et al., 2014; Zauderer et al., 2013; Berger et al., 2012; IR and radio polarimetry: Wiersema et al., 2012).  Aliu et al. (2011).
Aleksic et al. (2013) reported non-detections of Swift~J1644+57 using   MAGIC and VERITAS for energies greater 100 GeV.
%\LEt{ Single-sentence paragraphs are not allowed.} 
Evidence  for  quasi-periodical oscillations (QPOs) in the X-ray power spectrum of Swift~J1644+57   
with a period of  $\sim$ 200 s in  X-rays was suggested by  Reis et al. (2012) and  Saxton et al. (2012). In contrast to the  long-term  X-ray  lightcurve, the  radio emission of Swift~J1644+57 
continued  to  rise  (Zauderer  et  al., 2013; Berger  et  al.,  2012).
% see also our Figure~\ref{radio}). 

The $Swift$ X-ray lightcurve of Swift~J1644+57   decreased and  then (at 507th  day from the discovery) showed a sudden drop by a factor of $\sim$ 200, that is, it was no longer detectable with $Swift$ but only detectable by $Chandra$ (Zauderer et al., 2013) and XMM-Newton in longer time pointings  (Gonzalez-Rodriguez  et  al.  2014). A second event very similar to Swift~J1644+57 was discovered by the $Swift$/BAT % Burst Alert Telescope (BAT, 15 -- 150 keV, Barthelmy et al. 2005) 
on May 27, 2011  (Krimm et al. 2011)  from  Swift~J2058+0516, hereafter Swift~J2058+05. 
%by  Cenko et al. (2012).
This TDE source
%\LEt{ or 'these... sources are'.} 
is located at { a higher} redshift, 1.186 \citep{Komossa15,Cenko12}.

The evidence for a possible TDE  in Swift~J2058+05 % discovered by $Swift$ 
was reported by Cenko et al (2012) who showed that this source 
%Sw J2058+05 
demonstrated a luminous, long-term X-ray outburst with  a peak luminosity of $L_x \sim 3\times 10^{47}$ erg/s. 
The event was also accompanied by a strong radio emission \citep{Pasham15}. The associated host galaxy of  Swift~J2058+05 
%is observed at redshift, 1.185~\citep{Komossa15,Cenko12}, but 
was optically inactive. Because of the 
many similarities between Swift~2058+05 and Swift~J1644+57, Cenko et al. (2012) suggested that a similar outburst mechanism is 
consistent with multi-wavelength follow-up observations (Pasham et al. 2015).
% It is worth  noting that 
The X-ray  lightcurve of Swift~J2058+05 is similar to that of Swift~J1644+57 which shows an abrupt drop after  250 -- 300 days since the initial outburst. Because of 
rapid variability of the X-rays before the drop, an origin  was associated  with the vicinity of  SMBH rather than  with the forward shock location (Pasham et al. 2015).

The BH mass limits of  Swift~J2058+05 were derived  based on the X-ray turnoff 
($10^4 M_{\odot} \le M_{BH} \le 2\times 10^6 M_{\odot}$, 
%Pasham et al., 2015) 
as well as using the  BH mass upper limit estimate method of applying the X-ray variability timescale, $5\times 10^7$ M$_{\odot}$ (see Pasham et al.  2015). 
Furthermore, Pasham et al. (2015) assumed that the TDE source optical flux is dominated by the host galaxy and they constrained the BH mass of the central SMBH using the well-known bulge luminosity as a function of 
BH mass relations (e.g., Lauer et al. 2007). As a result they inferred the SMBH
mass of about $M_{BH} \le 3\times 10^7$ M$_{\odot}$.

 It is desirable to have an independent BH  identification  for its central %(compact) 
object  
%located  in the center of ESO~243-49 HLX-1 as an alternative to the dynamical method. 
as well as the BH mass determination  by an alternative to the aforementioned methods, 
based on luminosity and minimal variability time estimates only. 
A method of  the BH mass determination was developed by Shaposhnikov \& Titarchuk (2009), hereafter ST09, using a correlation scaling between X-ray spectral and timing (or mass accretion rate) properties observed for many Galactic BH binaries during  %during 
their spectral  state transitions.

he origin of the X-ray emission in TDE is still unclear. However, this is for Swift~J1644+57,
 there are two, well detected, prominent peaks in the  spectral energy distribution: one in the far-infrared and another   in the hard X-ray band (Bloom et al. 2011; Burrows et al. 2011). They can be modeled as a direct synchrotron emission  (single-component model) from radio to X-rays, with strong dust extinction in the optical and ultraviolet
%\LEt{ and? or? Please replace the slash as it is ambiguous.} 
bands.
 Alternatively, the radio and IR
%\LEt{ and? or? Please replace slash.} 
peaks are related to the  synchrotron emission and the X-ray peak is due to inverse Compton scattering of external photons, most likely disk photons (two-component blazar model). Another  possibility is that the X-ray emission is due to the inverse Compton emission at the base of the jet, while radio and IR
%\LEt{ replace slash.} 
synchrotron emission comes from the forward shock at the interface between the head of the jet and the interstellar medium (Bloom et al. 2011).

%Another interpretation of the high-energy (E > 100 keV) emission is that the
%spectra are formed due to Comptonization of soft (disk) photons by the bulk motion of matter falling
%into a black hole (Laurent \& Titarchuk 1999).  Niedźwiecki \& Zdziarski
%(2006) have argued that the non-detection of spectral breaks at $E < 500$ keV is
%contrary to the prediction of the Bulk Motion Comptonization (BMC) model. 
 %, one of the goals 
The goal of the present study is to investigate 
the { peak in }
X-ray observational data from Swift~J1644+57 and  Swift~J2058+05  and 
to infer their fundamental observational characteristics.
In particular, the soft X-ray spectrum of Swift~J1644+57 was fitted by a simple absorbed power law, although more complex spectral models were  also discussed (Burrows et al. 2011; Saxton et al. 2012). An average photon index during the first year % $2\times 10^5$ s
was $\Gamma \sim 1.8$ (Burrows et al. 2011; Levan et al. 2011). However, the physical meaning of this average value  has to be interpreted more carefully, as there were strong variations in hardness and  in the photon index between flares, with a change between $\Gamma\approx$ 1.3 and 3 (Levan et al. 2011). In particular, the photon index was harder when the source was brighter (Burrows et al. 2011; Kennea et al. 2011; Levan et al. 2011).

Our main goal in the present study  is the full analysis of the Swift/XRT follow-up data of Swift~J1644+57  from the beginning to the end of the 2011 -- 2012 outburst and Swift~J2058+05 during 2011 outburst decay. 
We  present spectral analysis and interpretation of  {\it RXTE}, $Swift$ and $Suzaku$ data of the sources. 
In \S 2 we show  the list of observations used in our  analysis, while 
in \S 3 we provide  details of  X-ray spectral fittings.  We discuss an evolution of 
the X-ray spectral properties during the high-low state  transitions in \S 4
and demonstrate the results of the scaling analysis, in order to estimate BH masses of Swift~J1644+57 and  Swift~J2058+05 in \S 5. % and \S 5.  
We  make our conclusions   in \S 6. % and \S 7. 

\section{Observations and data reduction \label{data}}

Swift~J1644+57 was  observed by   $Suzaku$ during April 6, 2011  and  May 17, 2012 (see \S~\ref{suzaku data}) along with the short-term {\it RXTE} observations on  March 30 -- 31,  2011 which we describe in \S~\ref{rxte data} and by $Swift$ 
during the period of 2011 -- 2016 (see \S~\ref{swift data}). 
{The $RXTE$ data (three observations) are related to the peak burst phase and probe harder X-ray energies (3 -- 100 keV). In contrast, the well-exposed $Suzaku$ data are very advantageous in determining  low-energy photoelectric absorption, which is presumably not associated with the source directly. 

Short-term {\it RXTE} observations of of Swift~J2058+05 on June 1,  2011 are 
described in \S~\ref{rxte data}. We 
also used publicly available data  by the $Swift$ Observatory obtained from  May 27, 2011 to December 7, 2011 
(see \S~\ref{swift data}). 
{
We extracted these data from the HEASARC archives and found that the $Swift$ %these 
data  cover  the decay phase of X-ray outburst for  Swift~J1644+57 and Swift~2058+05, as well as %and 
partly, catch the peak burst interval. %wide range of X-ray luminosities. 
% We extracted these data from  the HEASARC archives and found that the $Swift$ %these 
%data  cover  the decay phase of X-ray outburst  and partly, catch the peak burst interval. %wide range of X-ray luminosities. 
%In contrast, %It is recognized that 
%the well-exposed $Suzaku$ data are very advantageous in determining  low-energy photoelectric absorption,  which is presumably not related to the source directly. 
%%The $RXTE$ data (one observation) are related  to the peak burst phase and generally probe harder X-ray energies (3 -- 30 keV).
}
A summary of the X-ray observations analyzed in this work is given in Tables~1$-$3.

\subsection{\it Suzaku \label{suzaku data}}

We studied the TDE source, Swift~J1644+57 using the {\it Suzaku} data,  for    April 6 of  2011 and May 17 of 2012 observations (see Table~1). These particular observations correspond to nine and 416 days after the source discovery. 
We used  {\tt HEASOFT software package} (version 6.13) and calibration database 
 (CALDB) released on  February 10,   2012   %and 2011 September 13  
for  XIS. % and HXD, respectively.
Since background is dominant in the lower energy band, we used photons in the 1 -- 10 keV (for XIS0, and 3) 
and 1 -- 7 keV (for XIS 1) energy bands. % XIS 0+3 and XIS 1, respectively
%  We applied the unfiltered event files for each of the operational XIS detectors (XIS0, 1 and 3) %; Koyama et al.2007) 
%using the latest {\tt HEASOFT software package} (version 6.13)  
%and 
 
The data reduction and spectral analysis are performed following 
%the Suzaku ABC guide (http://heasarc.gsfc.nasa.gov/docs/suzaku/analysis/abc/) using HEASOFT
%6.12 and the newest CALDB (calibration database).
  the {\it Suzaku} Data Reduction Guide\footnote{http://heasarc.gsfc.nasa.gov/docs/suzaku/analysis/}. %  using the latest HEASOFT software package (v6.13). % and 
 We obtained cleaned event files by re-running the {\it Suzaku} {\tt pipeline} implementing the
latest calibration database (CALDB) available since  January 20 of  2013,    and also applying the associated screening criteria files. 

%Thus, we got the BL Lac spectra  from the filtered XIS event data %list 
%taking a circular region, centered on the source, of radius 6{\tt '}. %, and the corresponding background spectra 
%from an annulus region with 30{\tt "} and 180{\tt "}  radii. 
%Using the {\it Beppo}SAX sample we considered the 
%background region to be in the vicinity  of the source   extraction region. 
We  extracted spectra %and lightcurves   
from the cleaned event files using XSELEC and we generated  responses  for each
detector utilizing  the XISRESP script with a medium resolution.
The spectral and response files for the front-illuminated
detectors (XIS0, 1 and 3) were combined using the FTOOL
ADDASCASPEC, after confirmation of  their consistency. 
In addition, we grouped the spectra in order to have a minimum of 20 counts per energy bin. 
We carried out  spectral fitting  applying XSPEC v12.7.1. 
 The energy ranges around of 1.75 and 2.23 keV are not  used for spectral fitting because of the known artificial structures in the XIS spectra around the Si and Au edges.
%Additionally, owing to 
%strong absorption, the spectrum of 4U~1700-37 has very limited statistics below 3 keV. 
 Therefore, for spectral fits we  chose  the 0.3 -- 10 keV  range  for the XISs 
(excluding 1.75 and 2.23 keV points).
% see Figure~\ref{spectrum_Sz_sw}). % and the 15 -- 70 keV  range for the PIN spectrum. 
%We fitted the spectra
%simultaneously with all parameters tied, except the relative
%instrument normalizations which were kept free. 
%-------------------------------------------------------------------777777
\subsection{\it RXTE \label{rxte data}}

We  analyzed three {\it RXTE} observations of Swift~J1644+57  made between July 1997 and January 2001  related to different spectral states of the source.  
In addition, we  analyzed  only one   {\it RXTE} observation of  Swift~J2058+05   made on %1$^{st}$ 
June 1, 2011 close to peak burst state of the source.
%available data from  the public archive \citep{bradt93}. 
%In total, this set includes 425 observations taken at
%different states of 4U~1630--47. %the source.

Standard tasks of the LHEASOFT/FTOOLS 5.3 software package were %applied 
%used 
applied for data processing. For spectral analysis we %explored 
used PCA {\it Standard 2} mode data, collected in the 3 -- 23~keV energy range, %applying 
using 
%the most recent release of 
PCA response calibration (ftool pcarmf v11.7).
%{\bf NS !!! either v11.7 should be used or "most recent" should be removed!!!}. 
% BEGIN SE
{
%The fitting was carried out using the standard XSPEC v 12.6.0 fitting package. 
}
% END SE
The standard dead time correction procedure was applied to the data. 
In order to construct the broadband spectra of  the data we also used HEXTE detectors in the case
of Swift~J1644+57. 
% have  been also used.
%Herewith, for
 The spectral analysis of  the data  in the 19 %20 
-- 150~keV energy range was implemented  in order 
to account for the uncertainties in the HEXTE response and 
background determination.
We  subtracted a background corrected  in  off-source observations. 
%In order to account for the uncertainties in the HEXTE response and 
%background determination, only data  in the 19 %20 
%-- 200~keV energy range were 
%used for the spectral analysis. %The HEXTE data have been re-normalized based on the PCA.
 The data are available through the GSFC public archive,  
http://heasarc.gsfc.nasa.gov. Systematic error of 0.5\% was applied to all analyzed {\it RXTE}  spectra. 

In  Table~2
%-- \ref{tab:list_RXTE_20}  
we listed the  groups  of {\it RXTE} observations of  Swift~J1644+57 and Swift~J2058+05 tracing %covering 
thoroughly %complete range of 
the source evolution during different spectral states. %transition events. 
%On the whole %Thus 
%We have  made an analysis of {\it RXTE} observations  of BL Lac  during four years for three intervals 
%indicated by  blue rectangles in Figure~\ref{asm_1630} ($top$).
We modeled the {\it RXTE} %PCA 
energy spectra %were modeled 
using XSPEC astrophysical fitting software. 

\subsection{{\it Swift}  \label{swift data}}

The observational set  of Swift~J1644+57 is  extensive. The source was monitored by the XRT several times a day,
every day since  March 28, 2011.  We used the {\it Swift} observations  carried out from 2011 to 2016. 
%We performed an analysis of all $Swift$/XRT observations of M101 ULX-1 carried out from 2006 to 2013. 
The log of the {\it Swift}/XRT observations used in this paper  is shown in Table~3.
 We analyzed only photon-counting mode (PC) data. %The {\it Swift} source count rates never exceed 0.02 count s$^{-1}$, therefore  only photon-counting mode (PC) events 
%(selected in grades 0$-$12) were considered. 
The $Swift$-XRT/PC data (ObsIDs, shown in the second column of %the upper part 
Table~3
%\ref{tab:list_Swift})  
were processed using the HEASOFT v6.14, the tool XRTPIPELINE v0.12.84 and the
calibration files (CALDB version 4.1).
%We selected source events which were accumulated the grade 0$-$12 events.
The ancillary response files were created using XRTMKARF v0.6.0 and exposure maps generated by XRTEXPOMAP v0.2.7. 
We fitted the spectrum using the response file SWXPC\-0TO12S6$\_$20010101v012.RMF.
We also used the online XRT data product generator\footnote{http://www.swift.ac.uk/user\_objects/}  for independent check: 
light curves and spectra, including background and ancillary responce files 
%We also obtain  images %were extracted 
%using the online XRT data processing facility\footnote{http://www.swift.ac.uk/user\_objects/} 
%(5 http://www.swift.ac.uk/user objects/) 
(see Evans et al. 2007, 2009). 

The log of the {\it Swift}/XRT observations of Swift~J2058+05 used in this paper  is also shown in Table~3.
 %We analysed only photon-counting  (PC) mode data. %The {\it Swift} source count rates never exceed 0.02 count s$^{-1}$, therefore  only photon-counting mode (PC) events %(selected in grades 0$-$12) were considered. 
The $Swift$-XRT/PC data (ObsIDs) are  indicated  in the third column of %the upper part 
this table
%~\ref{tab:list_Swift})  
and they were  processed using the same software as that in the case of Swift~J1644+57.
% the HEA-SOFT v6.14, the tool XRTPIPELINE v0.12.84 and the
%calibration files (CALDB version 4.1).
%We selected source events which were accumulated the grade 0$-$12 events.
%The ancillary response files were created using XRTMKARF v0.6.0 and exposure maps generated by XRTEXPOMAP v0.2.7. 
%We fitted the spectrum using the response file SWXPC\-0TO12S6$\_$20010101v012.RMF.
%We also used the online XRT data product generator\footnote{http://www.swift.ac.uk/user\_objects/}  for independent check: 
%light curves and spectra, including background and ancillary response files 
We also obtained  images %were extracted 
using the online XRT data processing facility\footnote{http://www.swift.ac.uk/user\_objects/} 
%(5 http://www.swift.ac.uk/user objects/) 
(see Evans et al. 2007, 2009). 
%BEGIN NEW
%We grouped the $Swift$ spectra into four bands according to count rates (Sect.~3.1) 
%%: very high ("A"), high ("B"), medium ("C") and low ("D") count rates (see Fig.~2) 
%and fitted the combined  spectra of each band with the {\tt XSPEC} package (version 12.8.14).
In this way, we inspected the image in vicinity of Swift~2058+05
% (Figure~\ref{image}) 
to exlude the presence of other sources close to this source.
%TDE Swift J2058+05. 
For the Swift/XRT image 0.3 -- 10 keV of the source field of view  (FOV) we use 
%indicate 
the circle of 5 arcmin radius 
with the center of the source position [$\alpha = 20^h 58^m 19.76^s$, $\delta= +05^{\circ} 13' 29.8"$, J2000.0; \citet{Cenko12}]. 
The faint source is 
%presented 
at the bottom of the image (at 5.8 arcmin radius), but it is beyond the FOV used for our analysis. % 5 arcmin radius  from TDE Swift J2058+05
%(the faint source is more evident at 0.3-1 keV than at higher energies).
%Maybe we can say that no source contamination is present.
%BEGIN NEW
%We grouped the $Swift$ spectra into four bands according to count rates (Sect.~3.1) 
%%: very high ("A"), high ("B"), medium ("C") and low ("D") count rates (see Fig.~2) 
%and fitted the combined  spectra of each band with the {\tt XSPEC} package (version 12.8.14).

\section{Results \label{results}}

\subsection{X-Ray light curves \label{light curve}}

The 0.3 -- 10 keV  light curve of Swift~J1644+57 is shown in Fig.~\ref{lc} (top panel), for the time period  2011 -- 2012. 
%\LEt{ symbol colors etc should be described in the figure legen, and not repeated in the main text. 
%We agree!}
The source signal (with 2-$\sigma$ detection level) for PC/WT mode is indicated, as is 
the background level for PC/WT mode, respectively. 
We  found  %at least one global outburst of BL Lac peaked at MJD=56248 and intervals of moderate variability around 2012.   
%with a FRED (``fast rise exponential decay'') profile 
a complex count rate behavior at  the burst peak, which is  variable from  0.3 to 100 counts/s (in PC mode) within 100 s, and  
a long %\LEt{ please check I have not changed your intended meaning. Agree!} 
slow decay (from 1 to 0.1 counts/s) with rough duration of 500 days. 
For the rest of the {\it Swift} observations the source remained in the {low state}. 
%We  should note that individual {\it Swift}/XRT observations  in PC ({Photon counting}) 
%mode do not have enough counts in order to make  statistically significant spectral fits. 
Specifically, on   508th  day  (from the first detection on March 28, 2011 ) the source emission suddenly dropped below the Swift detection limit  (Sbarufatti et al. 2012).
%The  X-ray spectra of Swift~J2058+05 were obtained using the $Swift$/XRT  observations in 2011 during transition from the high/soft state (HSS) through the intermediate state (IS)  to the low/hard state (LHS). 
{ The light curve of Swift~J2058+05 in the 0.3 -- 10 keV range {using $Swift$/XRT} during the source outburs decay on  from May 27 to July 11,   2011 is presented in Fig.~\ref{lc_20}. }

\subsection{Hardness-intensity diagrams and {HR} curves  \label{HID_lc}}

%Before we proceed with %} %going to   
%details  of spectral fitting  we  %study 
%investigate a  so {called} hardness ratio (HR). 
%in order to quantify and characterize 
%the source spectrum. 
In application to the {\it Swift} data  of Swift J1644+57 we have defined the hardness ratio (HR) as a ratio of the hard and soft counts  in the 1.5 -- 10 keV and %$S$  and $H$ 
 0.3 -- 1.5 keV bands, respectively.
% However, at low counts, the posterior distribution of the counts ratio, $R$, tends to be skewed
%because of the Poissonian nature of data. Therefore we used the color, $C=log_{10}(S/H)$, which  a log 
%transformation of $R$, which provides the skewed distribution more symmetric (see e.g., Park et al. 2006).
%monotonic function of the counts $S$ and $H$ in 
%the soft (0.3 -- 1.5 keV) and hard (1.5 -- 10 keV) bands, respectively. Specifically,: $C=log_{10}\frac{S}{H}$. 
The HR value  strongly depends on  %evaluated %
 calculation of  the background counts.
% and instrumental effective areas. 

In the bottom panel of Fig.~\ref{lc} we show the HR curve  versus MJD time %1.51 -- 10 keV/0.3 -- 1.5 keV) 
using PC mode data of Swift~J1644+57. 
At the beginning, this curve shows variations tracking the light curve local flares, for which  the hardness rises when the average rate increases, and then saturation  when the average rate decreases. 
%(see $bottom$ panel of Fig.~\ref{lc}).
% But later phases with different trends can be identified. 
We observe a hardness plateau phase from  $\sim$ 2.5 to 16 days after the source discovery. 
%corresponding to a phase of average rate decrease (from $\sim$ 2.5 to 6 days) followed by a rapid rise by a factor
%of about 10 within $\sim$ 0.4 days and a shallow decay later.
% Then, we observe the saturation of the hardness ratio while the average rate in the light curve steadily decreases. 
 %we observe a relatively rapid ($\sim$ 1 day long) decreasing (i.e. softening) in the hardness 
%with a subsequent steep rising (i.e. hardening) phase that continues up to $\sim$ 100 days, when a final hardness plateau  phase begins.

It is interesting that we detect a sharp hardening of X-ray emission flux of Swift~J1644+57 near the  initial outburst part, which is typical for known Galactic BHs (GBHs) and usually  accompanied by the jet turn-on  [e.g., in GRS~1915+105, see \cite{Bel_Motta16}; % Belloni et el. 2008???, 
\cite{mf06}; \cite{tsei09}]. %Migliari et al. 2006, TS09).
%Titarchuk \& Seifina, 2009).

In Fig.~\ref{HID} we present  %demonstrates 
the  hardness-intensity diagram (HID), which demonstrates %  and thus, we show  
that different count-rate observations  are assocated with %correspond 
 different color regimes. The HR larger values correspond to  harder spectra.
% A hysteresis-like evolution is discernible in the color-intensity
%track in its %intermediate phase 
%middle part, which is caused by the differences between the rise and decay 
%tracks. % parameters. 
%Larger values of the {HR} indicate a , and vice versa. 
%Note, that 
% we have applied
A Bayesian approach was used  to estimate   the HR values and their errors [see \cite{Park06}]\footnote{A Fortran and C-based program which calculates the 
%hardness 
ratios using the methods described by 
% the paper of 
\cite{Park06} 
%which  is available for download from 
(see http://hea-www.harvard.edu/AstroStat/BEHR/)}. 
For clarity, we plot only one point with error bars (in the bottom right corner) to demonstrate 
typical uncertainties  for the count rate and HR.

Figure~\ref{HID} clearly shows %clear indicates 
that the {HR} monotonically drops %reduces  
with the soft %total count rate (0.3 -- 10 keV). %flux $S$ and achieves  
count rate (0.3 -- 1.5 keV). 
This particular sample  is similar to those of  most of outbursts of Galactic %low mass 
X-ray binary transients (see Homan et al. 2001; Belloni et al. 2006; 
Shaposhnikov \& Titarchuk, 2006;  ST09; TS09; Shrader et al. 2010; Munoz-Darias et al. 2014).

\subsection{X-Ray spectral analysis \label{spectral analysis}}

A number of  spectral models was used  in order to test them  for all available data sets for Swift~J1644+57 and Swift~J2058+05. Our goal is  to establish  an evolution between the low hard
%\LEt{ Please replace the slash, or define this use as an abbreviation.} 
and high soft
%\LEt{ See Note 14.} 
states based on spectral modeling.

\subsubsection{Swift~J1644+57}
The  X-ray spectra of Swift~J1644+57 obtained using the $Suzaku$/XIS  observations  in 2011 
(the high soft state) and 2012 (the low hard state).
In Fig.~\ref{spectrum_Sz_sw} we show examples of these spectra along with that obtained by the {\it Swift}/XRT. Two $EF_{E}$ {\it Suzaku} spectral diagrams related to the high soft
state and the low hard state are presented  in Fig. \ref{spectrum_Suz_70_90}.
While the spectrum detected   by  {\it RXTE}  and the combined {\it Swift-RXTE} spectral diagram are shown in Fig.~\ref{RXTE_spectrum_CompTB_E_cut_70keV} and Fig. 7 in Seifina et al. (2017, see the incoming A\&A paper).   %\ref{spectrum_Sw+RXTE_E_cut}, 
respectively.

%Figure~\ref{sw_spectrum} shows the best-fit spectrum during IS-LHS  of the decay transition in $EF_E$ units (top panel) and with $\Delta\chi$ (bottom panel) for Swift~J2058+05.

%A number of  spectral models was used  in order to test them  for all available data sets for Swift~J1644+57 and Swift~J2058+05. Our goal is  to establish  an evolution between the low/hard and high/soft states based on spectral modelling.
% in frame of the same model. 
We investigated the {\it Suzaku} %, {\it $Beppo$SAX, ASCA, RXTE}, and combined {\it Swift} spectra 
%related to  different spectral states (the HR bands, see Fig.~\ref{HID}) 
spectra for Swift~J1644+57  to check  the following XSPEC  spectral models: 
powerlaw, blackbody, the bulk motion Comptonization (BMC)  and their possible combinations modified by an absorption  model.
%\LEt{ single-sentence paragraphs are not allowed. Please connect this to the previous or following paragraph. Done!} 
We find that the absorbed  power-law model ($phabs*zphabs*zpowerlw$)  reveals a statistically-unacceptable fits with  the photon index $\Gamma_{pow}\sim$ 1.9 and the reduced chi-squared  $> 2$ ($Sz1$, see Tables~4-\ref{tab:list_SAX_Suzaku}). A similar result was   also previously reported by Burrows et al. (2011) and Reis et al. (2012).  
Moreover,  the thermal model ($zbbody$) gives us even worse fits. 
%than the power-law model. %However, the intermediate state spectra  (D-spectra) % for $Swift$ data)
% and $Suzaku$ data) spectra 
%cannot be fitted by any single-component model. Indeed, a simple power-law model produces a soft excess.  
%These significant positive residuals at low energies, less than 1 keV, suggest the presence of additional emission  components in the spectrum. 
As a result  we   checked a combination of $zbbody$ and $zpowerlw$  models. In this case the model parameters  
are  $N_{H, z=0.354}=(1.2 - 1.9) \times 10^{22}$ cm$^{-2}$; $kT_{zbb}=0.24-0.3$ keV and $\Gamma=1.3 - 1.9$ (see more detail
in Table~4). The best fits of the {\it Suzaku} spectra has been found  using   the   
 {BMC XSPEC} model \citep{tl97},   
% the {\it Bulk Motion Comptonization} 
%(BMC) model 
for which $\Gamma$ ranges from 1.1 to 1.8  for all observations (see Table~4
%\ref{tab:par_suz} % 2 
and Figs.~\ref{spectrum_Sz_sw}--\ref{spectrum_Suz_70_90}).  In Fig. \ref{spectrum_Sz_sw} we also show  the simultaneous spectra of Swift~J1644+57 obtained using  the $Suzaku$ and $Swift$/XRT %observation  
  observations. % ($Sz1$). 

Thus, we decided to analyze all available spectral data of Swift~J1644+57  using the {\tt XSPEC} BMC model.
%{ which a brief model description is presented below, for completeness.}
% For the sake of completeness, now we provide a short description of the $bmc$ model. % in \S 2.5 (????).
The BMC model uses a convolution of  a seed  blackbody  with an upscattering  Green's function, presented  in the framework of the BMC as a broken power law in which left and right wings have the spectral  indices $\alpha+3$ and 
$\alpha$, respectively [see a description of the BMC Comptonization Green's function in Titarchuk  \& Zannias (1998)  (TZ98) and compare with Sunyaev \& Titarchuk (1980)]. The BMC model  has as  the main parameters, $\alpha$, $A$ (related to the illumination fraction $f=A/(1+A)$), 
the seed blackbody (BB) temperature $T_s,$ and the BB normalization which is proportional to the seed BB luminosity 
and inversely proportional to $d^2$ where $d$ is a distance to the source   [see also Titarchuk \& Seifina (2016a), 
hereafter TS16a]. %\ref{geometry}).

%%%%% 1
{

%Although %not very important as 
%we are dealing with nearby object 
The TDE source,  Swift~J1644+57 is located at $z=0.354$ and thus, we should  
 take into account  the cosmological reddening of the spectrum due to the redshifted energy band 
[$E\to E(1+z)$]. In order to make this redding corrections  we used the energies {\tt XSPEC} command, which extends the maximum energy over which the model is being calculated to $(1+z)$ times the maximum energy in the response.  
As a result the BMC$^z$ model describes the outgoing spectrum as a convolution 
of the input seed blackbody-like spectrum, of which normalization is $N_{BMC}$ and color temperature is $kT_s$,  with the  Comptonization Green's function.
%Similarly to the ordinary {\it bbody} XSPEC model, bolometrical luminosity

%\begin{equation}
%L^z_{bol}=\int_0^{\infty} E(1+z)\times A^z(E) dE,
%\end{equation}
%where 
%$A^z(E)$ is the photon flux density of blackbody radiation 
%\begin{equation}
%A^z(E)=8.0525 \times K \times \frac{E(1+z)^2}{(1+z)(kT_s)^4}\times\biggl[\exp{\frac{E(1+z)}{kT_s}}-1\biggr]^{-1},
%\end{equation}
%and 
%kT - color temperature in keV (par1 for both COMPTB and BMC models), and
%$K=N^z_{BMC}$ is the normalization of the seed blackbody photon spectrum, defined as a ratio of 
%the source (disk) luminosity to the square of the distance
%\begin{equation}
%N^z_{BMC}=\biggl(\frac{L}{10^{39}\mathrm{erg/s}}\biggr)\biggl(\frac{10\,\mathrm{kpc}}{d_{10}(1+z)}\biggr)^2,
%\label{bmc_norm}
%\end{equation}
%\noindent where 
%$L_{39}$ is the source luminosity in units of $10^{39}$ ergs$^{-1}$, 
%$d_{10}$ is the distance to the source in units of 10 kpc and $z$ is fixed redshift parameter. 
}

%%%%%11
{In addition to the Galactic absorption, $N_{H, Gal}=2\times 10^{20}$ cm$^{-2}$ 
in this sky direction we also apply a multiplicative 
{$zphabs$ model with a column density, $N_{H, z=0,354}$ of $ 1.5\times 10^{22}$ cm$^{-2}$ 
in the local (comoving) frame at redshift of 0.354
%\LEt{ Connect this to the previous paragraph. Done!}. 
} As a result we obtained
 the best-fit results using the same model for the $Suzaku$ and $Swift$ 
 spectral data of Swift~J1644+57 in the 0.3 -- 10 keV energy range 
throughout all (high soft and low hard) states. However, to fit the $RXTE$ spectral data of Swift~J1644+57 in a wide energy range we have to apply a multiplicative component $zhighect$ to the $BMC$ model in order to  correctly reproduce
the high energy part of the source spectrum (see Table~\ref{tab:par_rxte} and Fig.~\ref{RXTE_spectrum_CompTB_E_cut_70keV}). 
For the cross-check, we combined $Swift$ and $RXTE$ spectra both obtained on March 31, 2011  and 
apply the Comptonization  COMPTB model  (see Fig. 7, in Seifina et al. 2017, A\&A) 
%incoming~\ref{spectrum_Sw+RXTE_E_cut}) 
which  is 
the XSPEC %Contributed 
model\footnote{http://heasarc.gsfc.nasa.gov/docs/software/lheasoft/xanadu/xspec/ models/comptb.html},
see \cite{F08}, for the direct determination of the high energy cutoff of the spectrum, $E_{cut}$ which is related to  the plasma temperature $kT_e$ for the thermal Comptonization ($E_{cut}\sim 2kT_e$). 
 The fit parameters of these models are  shown in Table~\ref{tab:par_rxte}.

Using the same model, we made the spectral analysis of $Suzaku$, $Swift$ and $RXTE$ observations 
and found that Swift~J1644+57 was in the three spectral states (the low hard,  intermediate and soft states). The best-fit 
$\Gamma$   are presented in Tables~4-\ref{tab:par_rxte}  (see also 
Figs. \ref{spectrum_Sz_sw}$-$\ref{spectrum_Suz_70_90}). Evolution of the source %transitions 
between the low state and high state is accompanied by 
%strongly correlates with 
a monotonic increase of the normalization parameter, $N_{BMC}$ %of the Compton component  
from 0.5 to 300$\times L_{33}/d_{10}^2$ erg/s/kpc$^2$ which correlates with an increase of  $\Gamma$ 
%when it  monotonically increases 
from 1.1 to 1.8 (see  Fig.~\ref{saturation}). It is also interesting that this source demonstrates monotonical 
growth of the radio flux density (15.4 GHz) along with the drop of X-ray brightness 
(see second panel from the top in Fig.~9).
%\ref{radio}). 

The  rapid  decline {during 2011 -- 2012 outburst decay}  is not seen in the radio 
%(Fig.~10)
%\ref{radio}), 
implying that X-ray and radio emission at least have  different  origins  at  later  times.  In fact, Zauderer  et al. (2013) interpreted the sudden drop in X-ray  emission as evidence for a change in accretion regime.  
Specifically, %Although,   
Zauderer  et al. related  turning off the jet production, with   the X-ray fainter  stage. 
 They also argued that the continued radio emission  emananting  from the forward shock related to the jet.

\subsubsection{Swift J2058+05}
For Swift~J2058+05  we also  investigated the {\it Swift} and {\it RXTE} %, and combined {\it Swift} spectra 
%related to  different spectral states (the HR bands, see Fig.~\ref{HID}) 
spectra and  checked  the  XSPEC  spectral models which were 
powerlaw, blackbody, the BMC,  and their possible combinations modified by an absorption model and redshift correction. 

We show examples of the $Swift$/XRT spectra in $EF_E$ and 
normalized count units (see Figs.~\ref{sw_spectrum}-\ref{three_sw_sp}). In particular, Fig.~\ref{sw_spectrum} shows the best-fit spectrum during the  intermediate-low hard state (IS-LHS)  decay transition in $EF_E$ units (top panel) and  $\Delta\chi$ (bottom panel). %for Swift~J2058+05. 
Figure~\ref{three_sw_sp} demonstrates the spectral evolution using  three representative $Swift$/XRT spectra in normalized count units (top panel) with $\Delta\chi$ (bottom panel) for the LHS, the IS and the high soft state (HSS). %spectral states 
%of Swift~J2058+05. 
%The spectrum detected by  {\it RXTE}  %and 
%combined with $Swift$/XRT is shown in 
%Figure~\ref{spectrum_Sw+RXTE_E_cut_20}.  

We also  found, as that for    Swift~J1644+57,   the absorbed  power-law model ($phabs*zphabs*zpowerlw$)  reveals a statistically-unacceptable fits 
%with  the photon index $\Gamma_{pow}\sim$ 1.7 and the reduced chi-squared  > 2 
(see Table~6
%\ref{tab:par_sw_20} 
for id=00032004001). 
A similar result was   also previously reported by \cite{Cenko12} and \cite{Pasham15}.  
Moreover,  the thermal model ($zbbody$) gives us even worse fits. 

We then   checked a combination of $zbbody$ and $zpowerlw$  models. In this case the model parameters  
are  $N_{H, z=1.185}=(1.9 - 2.3) \times 10^{21}$ cm$^{-2}$; $kT_{zbb}=1.2-5.7$ keV and $\Gamma=1.3 - 1.9$ (see more details 
in Table~6). However, the best fits of the {\it Swift} spectra has been found  using   the   
 {BMC XSPEC} model \citep{tl97},   
% the {\it Bulk Motion Comptonization} 
%(BMC) model 
for which $\Gamma$ also ranges from {1.2} to 1.8  for all observations similar to that  for Swift~J1644+57 (see Fig.~\ref{saturation}).
Thus, we decided to analyze all available spectral data of Swift~J2058+05  using the {\tt XSPEC} BMC model.

The TDE source,  Swift~J2058+05 is located at $z=1.185$ and thus, we should also  
 take into account  the cosmological reddening of the spectrum due to the redshifted energy band 
[$E\to E(1+z)$], see details in \S 3.3.1. We used the BMC$^z$ model to describe the outgoing spectrum.  

In addition to the Galactic absorption, $N_{H, Gal}=6.5\times 10^{20}$ cm$^{-2}$ 
in this sky direction we also apply a multiplicative 
{$zphabs$ model with a column density, $N_{H, z=1.185}$ of $ 2.5\times 10^{21}$ cm$^{-2}$ 
in the local (comoving) frame at redshift 1.185. We obtained
 the best-fit results using the same model for the $Swift$ and {\it RXTE} spectral data of Swift~J2058 in the 0.3 -- 10 keV and 2 -- 30 keV energy ranges, 
respectively, throughout all spectral states (HSS, IS and LHS). However, to fit the $RXTE$ spectral data of Swift~J2058+05 in a wide 
energy range we should also  apply (as for the case  of Swift~J1644+57)  a multiplicative component $zhighect$ to the $BMC$ model in order to  correctly reproduce the high energy part of the source spectrum 
[see Table~\ref{tab:par_rxte} (lower part) and Fig. 7 in Seifina et al. (2017) in the coming A\&A paper %Fig.~\ref{spectrum_Sw+RXTE_E_cut}, 
related to  the  case of  Swift~J1644+57]. 
For the cross-check,
as we did  for the  Swift J1644+57 data, 
we combined $Swift$ and $RXTE$ spectra, both obtained at the end of May, 2011,  and   apply the COMPTB model
%  (see Fig.~\ref{spectrum_Sw+RXTE_E_cut_20}),  
%which  is 
%the XSPEC model\footnote{http://heasarc.gsfc.nasa.gov/docs/software/lheasoft/xanadu/xspec/ models/comptb.html},
%see \cite{F08}, 
for the determination of  $E_{cut}$ which is
% related to  the plasma temperature $kT_e$ for the thermal Comptonization 
$E_{cut}\sim 2kT_e$). 
 The fit parameters of these models are  shown in Table~\ref{tab:par_rxte} (lower part).

\section{Discussion}
%\subsection{Swift~J1644+57}
%To answer the question can the TDE scenario be described in terms of (disk) accretion approach  in application to 
We carried out { a} % the 
detailed analysis of  the  Swift~J1644+57   spectra observed during outburst decay (2011 -- 2012) and %. 
% As a result,  we 
revealed  that  Swift~J1644+57 { shows a} %demonstrated the 
spectral evolution similar to that seen in Galactic BHs. Specifically,  based on $Swift$, $RXTE$ and $Suzaku$ observations we establish for the Swift~J1644+57 spectra  $\Gamma$ correlates 
with the  BMC normalization, $N_{BMC}$ (which is proportional to $\dot M$)
and {finally} saturates at high values of $\dot M$ (see Fig.~\ref{saturation}). %, where 
The  index, $\Gamma$, monotonically grows from 1.1 to 1.7 with $\dot M$ and then saturates  at 
$\Gamma_{sat}\sim1.8$ for  high values of $N_{BMC}$. % $\dot M$.
Previously, Titarchuk \& Zannias (1998) developed the semi-analytical theory of X-ray spectral formation in  the converging flow into a  BH. 
They argued %demonstrated 
that the spectral index of the emergent X-ray spectum saturated  at high values of the mass accretion rate %(at 
(higher than the Eddingtion one). Later analyzing  the {\it RXTE} data for many BH candidate sources  ST09, \cite{tsei09}, \cite{seit10}, and STS14 demonstrated  that this  index saturation effect 
{ is} observed
% to that established 
in many  Galactic BHs (see e.g., GRS 1915+105~\citep{tsei09}, GX 339--4, GRO J1655--40, 4U 1543--47, H~1743--322,  Cyg X--1,  XTE~J1550--564~(ST09)
%\citep{st09}, 
SS~433~\citep{tsei09}).  Previously, this scaling method was effectively applied  to estimate BH masses of %Galactic (e.g. ST09, STS13) and 
extragalactic black holes (TS16a; Titarchuk \& Seifina 2016b, hereafter TS16b; Sobolewska \& Papadakis 2009; Giacche et al. 2014; Titarchuk \& Seifina 2017, hereafter TS17). 
Recently the scaling method was successfully implemented  to estimate BH masses of  two ultraluminous X-ray 
sources,  ESO~243--49 HLX--1 (TS16a) and  M101 ULX--1 (TS16b). These findings suggest  that BH masses are of  the order of $10^4$ solar masses in these objects.
%\LEt{ please check I have retained your intended meaning. Correct!}. 

The levels of the {saturation} index  are  different and   presumably depend on the plasma temperature of the converging flow [see Monte Carlo  simulations by Laurent \& Titarchuk (1999), (2011)].  
For Swift~J1644+57 we  establish that the photon index saturates at the relatively low level, $\Gamma_{sat}$ around 1.7 -- 1.8 (see Fig.~\ref{saturation}).
% vs mass accretion rate $\dot M$. 
This low $\Gamma_{sat}$ level can { be related} %indicate 
to the high electron temperature of irradiating plasma $kT_e\sim$ 40 keV, 
which also is in agreement with the high cutoff energy $E_{cut}\sim$ 80 keV detected in the source spectra observed by $RXTE$ [see Figs.~\ref{RXTE_spectrum_CompTB_E_cut_70keV}  and  Fig. 7 in Seifina et al. (2017) in the incoming A\&A] 
%\ref{spectrum_Sw+RXTE_E_cut}). 

This relatively low $\Gamma_{sat}$  and consequently, the high $T_e$ value are consistent  with each other. %which is presumably formed around ``invisible'' central black hole. 
It is probable that a large  fraction of the mass accretion flow goes out of the disk in the subKeplerian manner.   That leads  to the strong  upscattering  of the disk photons in the Compton cloud (transition layer) located between the Keplerian accretion disk and the particle last stable orbit ($\sim 3R_{sch}$).
The geometry of the illumination of the Compton cloud by the soft (disk) photons is shown in TS16a.  
%    less effective X-ray emission than that in case of disk accretion regime. 
%On the other hand, such ``off-disk'' 
%mass accretion inflow onto central black hole can cause subsequent outflow as jet.
We note that  Very Large Array and Very  Long Baseline Array radio observations of this X-ray source point to possible outflow (Zauderer et al. 2011, see also Fig.~8). Using the  index-$\dot M$ correlation found in Swift~J1644+57 
%{and Swift~J2058+05}, 
%{\bf Please exclude this source, because here we discuss Swift~J1644+57 only. } \}
we can  estimate a BH mass  in 
this source  applying scaling of this correlation with those   detected  
in particular GBH sources. %and ULXs M101 ULX--1 and ESO~243-49 HLX--1 
%(see the details below). % 4.3).

%Using the  index-$\dot M$ correlation found in Swift~J2058+05 we can  also estimate a BH mass  in 
%this source  applying scaling of this correlation with those   detected  
%in particular GBH sources as well as in the TDE source, Swift~1644+57. 

\subsection{Black hole mass estimates \label{BH_mass}}
ST09  (see \S 3.2 there)   elaborated the scaling method in detail and derived  formulae for the BH determination. 
%In fact, sometimes the editors and the referees do not allow us to put once again the details if this technique in a new coming paper. 
 ST09  (see their Eqs. (4) { and (6)} ) showed that the disk (soft photon) luminosity is proportional 
%to a BH mass and the mass accretion rate and  using this equation ST09 showed in Eq. (6) that the luminosity is proportional 
to a dimensionless BH mass and a mass accretion rate in terms of the solar mass and the critical mass accretion rate, respectively. Moreover,  the normalization of the soft photons in the Comptonization BMC model, $N_{BMC}$ is proportional to the disk (soft photon) luminosity and 
inversely proportional to $d^2$ (where $d$ is the distance to a  particular source).  
Hence, one can estimate a BH mass using this relation. The ratio of these normalizations for target and reference sources in the same spectral state is presented by Eq. (7) in ST09.

\subsubsection {Swift~J1644+57\label{BH_mass_1644}}
To estimate the BH mass, $M_{BH}$ of Swift~J1644+57, we chose the galactic BHs, GRO~J1655--40, 
Cyg~X--1, 4U~1543--47 and GX~339--4 
%XTE~J1550--564, H~1742--322 
(see ST09) as the reference sources 
% and 4U~1630-47 (see STS14)] and extragalactic source M101 ULX-1 (see TS16),  
for which the BH masses %, {$M_{1655}=6.3\pm 0.3$ M$_{\odot}$ (see Greene et al. 2001; Hjellming \& Rupen 1995) 
and  distances % {$d_{1655}=3.2\pm 0.2$ kpc  (see Jonker \& Nelemans 2004) 
were well established now 
(see Tables~7-8).   
%{\bf 
We note that the  BH masses of these BHs %  in  { GRO~J}1655--40 was  
were also estimated using  dynamical methods. 

For  a BH mass estimate  we  used $N_{BMC}$ of the reference sources and %.  
%and  index-QPO frequency  
%patterns previously  found  in ) 
%Thus, we 
  the index versus $N_{BMC}$  correlations for the reference sources, GRO~J1655--40, Cyg~X--1, 4U~1543--47 and GX~339--4,  
  with that of the target source Swift~J1644+57 (see Fig.~\ref{scaling}). 
The value of the  index saturation for the  Swift~J1644+57, $\Gamma_{sat}^{sw}\sim 1.78$,  
is close to $\Gamma_{sat}^{gro}\sim 1.98$ of GRO~J1655--40 
{ 
as well as $\Gamma_{sat}^{cyg}\sim 2.11$, $\Gamma_{sat}^{4u}\sim 2.15$ and $\Gamma_{sat}^{gx}\sim 2.24$ of  
Cyg~X--1, 4U~1543--47 and GX~339--4, respectively}.
% but is not the the same. 
While %But 
such a low index saturation level has never been detected in 
any Galactic or extragalactic BH source (see i.e., ST09 and TS17), %. However, 
the  slopes of $\Gamma$ -- $N_{BMC}$ correlation are the same   for both the target and reference sources (see Fig. \ref{scaling}).  
%Furthermore, as one can see from , the correlations of the target source (Swift~J1644+57) and the 
%reference source have  similar shapes. % and index saturation levels.
Hence,  it allows us to perform a reliable scaling of this reference source correlation with that of Swift~J1644+57. 
The scaling procedure was made in a similar way to that in  TS17; TS16a, b and ST09
% Titarchuk \& Seifina 2016b, hereafter TS16b. We used an analytical %{\bf fitting} 
where the authors applied an analytical  approximation  %generally based on the parameterization 
of the $\Gamma(N_{bmc})$ correlation, %{\bf approximated} 
determined by a function %(see also ST09)
 %according to ST09 

\begin{equation}
{\cal F}(x)= {\cal A} - ({\cal D}\cdot {\cal B})\ln\{\exp[(1.0 - (x/x_{tr})^{\beta})/{\cal D}] + 1\}
\label{scaling function}
\end{equation}
with $x=N_{bmc}$.

Fitting of the observed correlation by  this function ${\cal F}(x)$
provides us a number of the best-fit parameters $\cal A$, $\cal B$, $\cal D$, $x_{tr}$, and $\beta$
 More detailed description of these parameters is given in TS16a.

 One should 
rely on the same shape of the $\Gamma-N_{bmc}$ correlations for the target source and those for the reference sources in order to make  the BH mass determination for the target source using  the scaling method (ST09).  
{ A BH mass, %To estimate the BH mass,  
$M_t$  of Swift~J1644+57 (target source) can be estimated by %one should shift 
shifting the reference source correlation along $N_{bmc}-$axis  to that of the target source (see Fig.~\ref{scaling}) 
%and parameterization of the correlations for the reference and target sources in 
%Table \ref{tab:parametrization_scal}.

\begin{equation}
M_t=M_r \frac{N_t}{N_r} 
\left(\frac{d_t}{d_r}
\right)^2 f_G,  
%=  C_0 {N_t} {d_t}^2 f_G,
\label{scaling coefficient}
\end{equation}
\noindent where the subscripts ``$t$'' and ``$r$'' correspond to the ``target'' and ``reference'' sources, respectively
% r stands for the reference  sources  
and a  geometric factor,  
$f_G=(\cos\theta)_r/(\cos\theta)_t$, the inclination angles $\theta_r$,  
$\theta_t$ and $d_r$, $d_t$ are distances to the reference and target sources, respectively (see ST09), %Here subscripts $r$ and 
%The geometrical factor $f_G$ takes into account the accretion geometry, particularly for the disk geometry,  while $f_G\sim1$ in the case of  spherical accretion 
%%One can see values of $\theta$ in  
%%Table \ref{tab:par_scal}) and  if some of these $\theta$-values  are unavailable then we assume that %$\theta_t\sim \theta _r$ and correspondingly 
%%$f_G\sim1$. 
%$C_0=(1/d_r^2)(M_r/N_r)$ is the scaling coefficient, 
%for the pair (target and reference source), 
values $M_t$ and $M_r$ are in solar masses.

Figure~\ref{scaling} shows   the $\Gamma-N_{bmc}$ correlation  for Swift~J1644+57 using $Swift$, 
$Suzaku$, and {\it RXTE} spectra (see also Tables~4-\ref{tab:par_rxte}) 
along with the correlation for the GBHs, GRO~1655--40 , GX~339--4 , Cyg~X--1 , and 4U~1543--47.
%BH masses and distances for this target-reference pairs are presented  in Table~\ref{tab:par_scal}. %7. 

}
%%%% END of BF

%A BH mass, $M_t$  for Swift~J1644+57 can be estimated using  a relation (see TS16a)
%\begin{equation}
%M_t= C_0 {N_t} {d_t}^2 f_G %cos(\theta_r)/cos(\theta_ULX),
%\label{C0 coefficient}
%\end{equation}
%\noindent where 
%the scale coefficient for each scaling pair    %coefficient 
%$C_0=(1/d_r^2)(M_r/N_r)$ is the scaling coefficient, 
%%for the pair (target and reference source), 
%values $M_t$ and $M_r$ are in solar masses and $d_r$ is the distance to a particular reference source  measured in 10 kpc.

We use values of $M_r$, $d_r$, $d_t$, and $\cos (i)$ from Table~8
%\ref{tab:par_scal} % 1 
and then calculate the lowest limit of the mass, $M_t$  using the best fit value of  $N_t$ %$N_t= (6.59\pm 0.09)\times 10^{-6}$ 
taken them at the beginning of the index saturation   (Fig. \ref{scaling}) and measured
in units of $L_{39}/d^2_{10}$ erg s$^{-1}$ kpc$^{-2}$ [see Eq.~(1) and  Table 7
%\ref{tab:parametrization_scal}
 for values of the parameters of function ${\cal F}(N_t)$] .
%Using $d_r$, $M_r$, $N_r$ (see ST09) we found that  $C_0\sim 3.3$ for GRO~J1655--40. 
To determine the distance to Swift~J1644+57 we used  the formula (for $z < 1$)
\begin{equation}
d_{sw}=z_{sw}c/H_0
\label{bllac_distance}
,\end{equation} 
where  the redshift $z_{sw}=0.354$  for Swift~J1644+57 (see { Wright 2006}), % \cite{Urry00}], 
$H_0=70.8\pm 1.6$ km s$^{-1}$ Mpc$^{-1}$  is the Hubble constant and $c$ is the speed of light.  
Finally,  we obtained 
that $M_{sw}\ge 7.8\times 10^6~M_{\odot}$ 
($M_{sw}=M_t$) %, {for $N_H^h$ case}) and 
%$M_{ULX}\ge 1.5\times 10^4~M_{\odot}$ (for $N_H^h$ case)}, 
assuming $d_{sw1}\sim$1.5 Gpc. %~
%\citep{Urru00} 
%%and  $f_G\sim1$. 
Thus,  we obtained a lower limit to the mass due the unknown inclination.
We present all these results in Table~8.

%It is worth noting that  the inclination of Swift~1644+57 may be different from those  for the reference Galactic source 
%(e.g., $i\sim 70^{\circ}$ for GRO~J1655--40), therefore we take this  BH mass estimate for Swift~1644+57 as the 
%lowest BH mass value  because of  $M_{sw}$ is a reciprocal function of $\cos (i_{sw})$
% BEGIN NEW
%[which is followed from Eq.~\ref{C0 coefficient} with taking into account that $f_G=(\cos\theta)_r/(\cos\theta)_t$ there]. %In this context, we can suggest  that  the lower $N_H$ case corresponds to   

The obtained  BH mass estimate is in agreement with a ``fundamental plane'' estimate,  
$M_{sw}\sim 3\times 10^6 M_{\odot}$  (see Miller \& Gultekin, 2011). In addition, using minimum timescales and 
 the  variability method, Liang \& Liu (2003)  obtained an estimate of a BH mass in the interval,  $M_{sw}\sim 10^7 - 10^9 M_{\odot}$, 
{ which is consistent  with  our BH mass estimates using the scaling technique.}

We  derived the bolometric luminosity based on the normalization of the BMC model in the range of %BMC model  between 
%$2\times 10^{43}$ erg/s and $3\times 10^{45}$ erg/s
$10^{43}$ -- %erg/s and 
$10^{45}$ erg/s  (assuming isotropic radiation). The relatively high  isotropic bolometric luminosity 
of Swift~J1644+57 %in the 0.3 -- 10 keV band  
in the first  weeks after the initial outburst
is the main argument for a tidal disruption event  as an origin of this phenomenon because  the host galaxy  was not known as an AGN before the flare. %Furthermore,
The peak bolometric luminosity is in agreement with  the derived BH mass using our scaling technique. We  note that our peak bolometric luminosity $L^{peak}_{bol}$  for Swift~J1644+57 is  less than previous estimates of $L^{peak}_{bol}$ obtained by different authors (see for example, Komossa 2015; Saxton et al. 2012; Mangano et al. 2016). 
In particular, Komossa (2015) reported that $L^{peak}_{bol}\sim 10^{49}$ erg/s, which  
can point to  either  a larger BH mass ($M_{BH}>10^9$ M$_{\odot}$) or to  supercritical accretion regime  (with $\dot M > \dot M_{Edd}$). However, for $M_{BH}$ in the $5\times 10^6 - 10^7$ M$_{\odot}$ range, the disk is  expected to have a bolometric disk luminosity in the $(0.5-1)\times 10^{45}$ erg/s range which is less than the critical one,  $L_{Edd}$. Therefore, an application of the Comptonization model (BMC)  to the spectra of Swift~J1644+57 and our scaling method give us  the most reliable BH mass estimate.

{Indeed,  the inferred  luminosity is close to  the Eddington limit for
a $10^7$ solar masses BH, $L_{Edd} =4\pi c G M_{BH}/\sigma_{\rm T}\sim 1.4\times10^{45}$~erg/s.
%\begin{equation}
% L_{Edd} = \frac{4\pi c G M_{BH}}{\sigma_T}\sim 1.4\times 10^{45} ~~ {\rm erg/s}. 
%\label{ed_lum}
%\end{equation}
%for $M_{BH} = 10^7$ M$_{\odot}$). 
%suggesting that the emission can originate in the outflow, for example in a relativistic jet zone (see Fig.~9 with 
%the radio flux density evolution shown in the green point panel) rather than an accretion disk 
%(see also asymmetrical jet-like structure around the source shown in Fig.~1). 
The relatively high electron temperatures, $kT_e\sim$34 keV (see Table~5)  are obtained using $Comp$TB model  and the {\it RXTE} data for Swift~J1644+57 spectra   (see \S~3.4 and Fig.~\ref{RXTE_spectrum_CompTB_E_cut_70keV}).
%$\ref{spectrum_Sw+RXTE_E_cut}). 
For such  a high $T_e$ the effective scattering cross-section, $<\sigma>$ is less than the Thomson one, $\sigma_T$ 
[see, for example \cite{tss14}] and consequently, the critical luminosity is higher than the Eddington one.  But the temperature-corrected 
critical luminosity in the case of Swift~J1644+57 is  higher than  $L_{Edd}$ of $1.4\times10^{45}$ erg/s and thus, the observed luminosity 
from this TDE source is actually a few times less than the critical one. 

\subsubsection{Swift~J2058+05\label{BH_mass_12058}}
To estimate a BH mass
%$M_{BH}$ 
of Swift~J2058+05, we chose the galactic BH, GRO~J1655--40 (see ST09), {Cyg~X--1, GX~339--4, 4U~1543--47} and the TDE source, Swift~J1644+57 (see \S 4.1.1)  
as the reference sources for which the BH masses %, {$M_{1655}=6.3\pm 0.3$ M$_{\odot}$ (see Greene et al. 2001; Hjellming \& Rupen 1995) 
and  distances % {$d_{1655}=3.2\pm 0.2$ kpc  (see Jonker \& Nelemans 2004) 
were  established now 
(see Tables~7$-$8).   
We note that the  BH mass of  { GRO~J}1655--40 was  determined using  dynamical methods. 

For a BH mass estimate  we  used $N_{BMC}$ of the reference sources and the index versus $N_{BMC}$  correlations 
for the { abovementioned}  reference sources,  %GRO~J1655--40 and  Swift~J1644+57 
with that of the target source, Swift~J2058+05 (see Fig.~\ref{scaling}).
The value of the  index saturation for the Swift~J2058+05, $\Gamma_{sat}^{sw2}\sim 1.76$, is close to 
$\Gamma_{sat}^{sw1}\sim 1.78$ for  Swift~J1644+57,  as well as to $\Gamma_{sat}^{gro}\sim 1.98$ for GRO~J1655--40, 
{
as well as $\Gamma_{sat}^{cyg}\sim 2.11$, $\Gamma_{sat}^{4u}\sim 2.15$ and $\Gamma_{sat}^{gx}\sim 2.24$ of  
Cyg~X--1, 4U~1543--47 and GX~339--4, respectively}.
While such  low index saturation levels (in Swift~J2058+05 and Swift~J1644+57) have never been found in any Galactic 
or extragalactic BH source (see ST09 and  {\S~\ref{BH_mass_1644}}). %Titarchuk \& Seifina 2017, hereafter TS17 ).  

%123

The  slopes 
of $\Gamma$ -- $N_{BMC}$ correlation are the same for all the target and reference sources (see Fig. \ref{scaling}).  
Therefore,  it allows us to perform a reliable scaling of these reference source correlation with that of Swift~J2058+05. 
The scaling procedure was made in a similar way  as \S 4.1.2 and in  TS17, TS16a,  Titarchuk \& Seifina (2016b) (hereafter TS16b) and ST09 
% Titarchuk \& Seifina 2016b, hereafter TS16b. We used an analytical %{\bf fitting} 
where the authors applied an analytical  approximation  %generally based on the parameterization 
of the $\Gamma(N_{bmc})$ correlation (see Eq. \ref{scaling function}). 

A BH mass, %To estimate the BH mass,  
$M_t$  of Swift~J2058+05 (target source) can be estimated using  by the same method as that for Swift J1644+57. %one should shift 
We shift the reference source correlations along $N_{bmc}-$axis  to that of the target source (see Fig.~\ref{scaling} and Eq. (\ref{scaling coefficient}})). 
Figure~\ref{scaling} shows the $\Gamma-N_{bmc}$ correlation for Swift~J2058+05 using $Swift$ , 
and {\it RXTE} 
(see also Tables~5$-$6) 
along with the correlation for the GRO~1655--40  and  Swift~J1644+57 using 
$Swift$, $Suzaku$,  and {\it RXTE} spectra.  
%determined by a function 
We use values of $M_r$, $d_r$, $d_t$, and $\cos (i)$ from Table~8
%\ref{tab:par_scal} % 1 
and then calculate the lowest limit of the mass, $M_t$  using the best fit value of  $N_t$ %$N_t= (6.59\pm 0.09)\times 10^{-6}$ 
taken %them 
at the beginning of the index saturation  (see Fig. \ref{scaling}).
% and measured
%in units of $L_{39}/d^2_{10}$ erg s$^{-1}$ kpc$^{-2}$ [see Table \ref{tab:parametrization_scal_20}
% for values of the parameters of function ${\cal F}(N_t)$, see Eq.~\ref{scaling function}].

To determine the distance to Swift~J2058+05 we use  the formula (for $z > 1$)
\begin{equation}
d_{2058}\approx \frac{2 c}{H_0} 
\left[
1-\frac{1}{\sqrt{z_{2058}}}
\right], \\
\label{bllac_distance}
\end{equation} 
where  the redshift $z_{2058}=1.185$  for Swift~J2058+05 (Wright 2006}). %, % \cite{Urry00}], 
%$H_0=70.8\pm 1.6$ km s$^{-1}$ Mpc$^{-1}$  is the Hubble constant and $c$ is the speed of light.  
Thus,  we obtain that $M_{2058}\ge 2\times 10^7~M_{\odot}$ 
($M_{sw2}=M_t$) %, {for $N_H^h$ case}) and 
%$M_{ULX}\ge 1.5\times 10^4~M_{\odot}$ (for $N_H^h$ case)}, 
assuming $d_{sw2}\sim$3.7 Gpc. %~
%\citep{Urru00} 
%%and  $f_G\sim1$. 
Thus,  we estimated a lower limit of  the BH mass due to the unknown inclination.
We present all these results in Table~8}.

The obtained  BH mass estimate is in agreement with the mass limits derived  based on the X-ray turnoff 
($10^4 M_{\odot} \le M_{BH} \le 2\times 10^6 M_{\odot}$, Pasham et al., 2015) as well as using the  method of estimate of a
BH mass upper limit applying the X-ray variability timescale, $5\times 10^7$ M$_{\odot}$ (see Pasham et el.  2015). 
Furthermore, Pasham et al. (2015) assumed that the TDE source optical flux is dominated by the host galaxy and they 
constrained the BH mass of the central SMBH using the well-known bulge luminosity as a function of 
BH mass relations (e.g., Lauer et al. 2007). As a result they inferred the SMBH
mass of about $M_{BH} \sim3\times 10^7$ M$_{\odot}$, which is consistent  with  our BH mass estimates using the scaling technique.

We  derived the bolometric luminosity, based on the normalization of the BMC model, in the range of %BMC model  between 
%$2\times 10^{43}$ erg/s and $3\times 10^{45}$ erg/s
$10^{44}$ -- %erg/s and 
$10^{45}$ erg/s  (assuming isotropic radiation). The relatively high  isotropic bolometric luminosity 
of Swift~J2058+05 %in the 0.3 -- 10 keV band  
in the first  weeks after the initial outburst is the main argument for a TDE  as an origin of 
this phenomenon because  the host galaxy  was not known as an AGN before the flare. 
In particular, 
Pasham et al (2015) (see the top panel in their Fig.~1).
%\LEt{ please check I have not changed your intended meaning. Correct!}), 
and Komossa (2015) speculated that $L^{peak}_{bol}\sim 3\times 10^{47}$ erg/s, and they  estimated that  the  BH mass,  
$M_{BH}>10^7$ M$_{\odot}$ which is  similar to our scaling technique estimate
(our  inferred  luminosity at the maximum for Swift~J2058+05  is close to  the Eddington limit for
a $10^7$ solar masses BH, see \S 4.1.1).

The relatively low electron temperatures, $kT_e\sim$ 4 -- 10 keV (see also Table~\ref{tab:par_rxte}; lower part)  are obtained using CompTB model  and the {\it RXTE} 
data for Swift~J2058+05 spectra   (see \S~3.2.2)
% and Fig.~\ref{spectrum_Sw+RXTE_E_cut}). 
For such  a low $T_e$ the effective scattering cross-section, $<\sigma>$ is about of the Thomson one, $\sigma_T$ 
[see, for example \cite{tss14}] and consequently, the critical luminosity is close the Eddington one.

\subsection{Comparison of spectral and timing 
characteristics of TDE sources Swift~J2058+05 and Swift~J1644+57 \label{comparison}}

Because of the many similarities between Swift~J2058+08 and Swift~J1644+57, \cite{Cenko12} %Cenko et al. 
suggested a similar outburst
mechanism, consistent with multi-wavelength follow-up observations (Pasham et al. 2015). Below, we  present  a comparison of these sources in terms of spectral and timing properties to further reveal 
the similarities and the differences between TDE sources Swift~J2058+05 and Swift~J1644+57.

\subsubsection{Saturation levels of the photon index \label{satur_level}}

%and differences between Swift~J2058+05 and Swift~J1644+57 \label{BH_mass}}

%1) the same saturation level of the Photon Index; 

The TDE sources Swift~J2058+05 and Swift~J1644+57 demonstrate a similar behavior of the photon index 
versus mass accretion rate (or our BMC normalization). The saturations of the photon index 
occurs  at the same saturation level $\Gamma_{sat}\sim 1.8$ for both of these sources.

\subsubsection{Difference of the electron temperature  ranges in Swift~J2058+05 and Swift~J1644+57\label{satur_level}}

A comparison of the best-fit spectral parameters for these two TDE sources shows that the ranges of seed (disk)  temperatures are
similar for both of these objects, namely, $kT_s$ = 100 -- 400 eV (see  Tables 4, 6). On the other hand, values
of the electron temperature $kT_e$ are quite different. These values, $kT_e$ vary in a wide range 
up to $kT_e$ = 35 keV for Swift~J1644+57, while for Swift~J2058+05 $kT_e$ they are  between  4 keV and 10 keV.
% (see Table~\ref{tab:par_rxte} and Table 5 in TS17). 
The reason for this difference of temperature ranges is not so %quite 
obvious but we can suggest that the soft (disk) photon  emission is stronger in  Swift~J2058+05 than that in the case of Swift~J1644+57 and thus, $kT_e$ are lower in  Swift~J2058+05  than that in Swift~J1644+57.

\subsubsection{Similarity of timing characteristics during decay phases in Swift~J2058+05 and Swift~J1644+57 \label{radio1}}

Similarly to Swift~J1644+57, the lightcurve of Swift J\-2058+05 shows an abrupt drop during  250-300 days  (see, e.g., Cenko et al, 2012; Pasham et al. 2015) while secular decline is described by  different decay rates. More specifically, 
the decline in flux of Swift~J1644+57 is consistent with the $t^{-1.5-1.67}$~(Bloom et al. 2011; Levan et al. 2011; Mangano et al. 2016), while that for Swift~J2058+05 is consistent with $t^{-2.2}$~\citep{Cenko12}. The reason for this difference for flux decline rates is not clear up to now  but it can be associated with different outflow plasma conditions in these two sources. 
The count rate in Swift J2058+05 decreases by a factor of 150 {(see Fig.~\ref{lc_20})} which is comparable with that in Swift~J1644+57, a factor of $\sim$ 100 decline, (see Levan \& Tanvir 2012; Sbarufatti et al. 2012; Zauderer et al. 2013). 

Notably, in both of these sources, the X-ray  dimming occurs on a similar timescale after disruption.
In the case of Swift~J1644+57, Zauderer et al. (2013) argued that this sudden decrease in the flux is caused by  the drop  of an accretion flow from a super-Eddington to a sub-Eddington rates.
Using numerical simulations by Evans \& Kochanek (1989) and De Colle et al. (2012) one can support this scenario. 
%However, If one set that the 
%same process is responsible for  in Swift~J2058+05, one can try to estimate the mass of the black hole 
%by equating the luminosity at turnoff to the Eddington luminosity. 
We note that applying the abrupt flux change, Pasham et al. (2015) constrained the black 
hole mass $M_{BH}$ in the range of $10^4 M_{\odot}$ to $2\times 10^6 M_{\odot}$ in Swift~J2058+05.

\subsubsection{Comparison of spectral evolution as a function of the
 normalization for Swift~J2058+05 and Swift~J1644+57 \label{normalization}}

We can also compare spectral parameter evolution for Swift~J2058+05 and Swift~J1644+57 using the BMC normalization. We note that  the
distances to these sources are different (see Table 8). Specifically, for Swift~J2058+05 the distance is about 3.7 Mpc, whereas for Swift~J1644+57 
it is 1.5 Mpc. In Fig.~\ref{scaling}, we show  correlations of BMC normalization, presumably proportional to mass
accretion rate, and the photon index $\Gamma$ for these two TDE sources. Swift~J1644+57 demonstrates a wider range of BMC 
normalization with a longer saturation part (by a factor of two higher than that for Swift~J2058+05), while $kT_s$ are almost the  same for both sources (100 -- 400 eV).

\section{Conclusions}

A  stellar tidal disruption event (TDE)  presents a new chance to estimate the mass of accreting 
supermassive black holes. The flare events occur after disruption of a star's
orbit at  about  ten of Schwarzschild radii from the central supermassive
BH. % and as a result  the star is torn apart by the BH's tidal force. 
A large amount of gas is suddenly injected close to a BH event horizon,
%and moreover, we can also anticipate the launch of a relativistic jet 
as this stellar debris gets accreted [\cite{Giannios_Metzger11}; \cite{van Velzen11}].

We find the transition from the high to low states observed in Swift~J1644+57 and Swift~J2058+05 during decay outburst phase  using the set of $Swift$, $Suzaku$, and $RXTE$ observations. %(2008 -- 2015) and %, %observations during  2006 -- 2013, 
%$Suzaku$ (2006) and $Chandra$ (2000, 2004 -- 2005) observations. 
We reveal a validity of the fits of the observed spectra using the BMC model   
for all observations, independently of the spectral state of the source. 

We investigated the X-ray outburst properties of TDEs 
%Swift~J1644+57 
%and confirmed the presence of spectral state transitions 
during the outbursts using of hardness-intensity diagrams %(Godet et al. 2009; Servillat et al. 2011) 
and the index$-$normalization (or $\dot M$) correlation, which were similar to 
those in Galactic BHs. 
In particular, we find that Swift~J1644+57 approximately follows the $\Gamma-\dot M$ correlation previously obtained for 
%extragalactic IMBH source M101 ULX-1 and 
the Galactic BHs,  GRO~J1655--40,  GX~339--4, Cyg~X--1  and 4U~1543--47  %X~339--4 and 4U~1534--47 %4U~1630-472, XTE~J1550-564 and H~1743-322 
%with taking into account  
%particular values of the $M_{BH}/d^2$ ratio 
(see Fig.~\ref{scaling}).
The photon index of Swift~J1644+57 spectrum is  in the range of $\Gamma = 1.1 - 1.8$. 
We also find that Swift~J2058+05 approximately follows the $\Gamma-\dot M$ correlation  obtained for 
%extragalactic IMBH source M101 ULX-1 and 
the Galactic BH,  GRO~J1655--40,  and Swift~J1644+57. %GX~339--4, Cyg~X--1  and 4U~1543--47  %X~339--4 and 4U~1534--47 %4U~1630-472, XTE~J1550-564 and H~1743-322 
%with taking into account  
%particular values of the $M_{BH}/d^2$ ratio 
%(see Fig.~\ref{scaling}).
The photon index of the Swift~J2058+05 spectra
%and Swift~J1644+57  spectra 
is    in the range of $\Gamma = 1.2 - 1.8$.
 
We used the observed index-mass accretion rate correlation to estimate $M_{BH}$  in Swift~J1644+57 {and Swift~2058+05}. 
This scaling method was  successfully implemented  to find  BH masses of Galactic (e.g., ST09, STS14) 
and extragalactic black holes [TS16a,b; \cite{sp09}; \cite{Giacche14}; TS17].  
%We found  values of 
%BEGIN NEW
%{
%$M_{BH}\geq 7\times 10^6 M_{\odot}$.   
%}  
% The previous  BH mass ranges  $(0.2 - 100) \times 10^7$ M$_{\odot}$ [Levan et al. (2011); Graham (2012); Miller \& G$\ddot u$ltekin (2011); \cite{Pasham15,lauer07}] and $(0.001 - 5) \times 10^7$ M$_{\odot}$ [\cite{Pasham15,lauer07}] are
%for   Swift~J1644+57 {and Swift~2058+05, respectively, obtained 
  %evaluated {\bf for Swift~J1644+57} 
%using the alternative methods.
% are consistent with  
%{\bf and  estimated {\bf for Swift~J2058+05} with the alternative methods 
%[see \cite{Pasham15,lauer07}].
%} our BH estimates.
% with  the inferred  low temperatures of the seed (disk) photons $kT_s$  we argue  that 
%in X-ray spectrum of M101 ULX-1, 
% that M101 ULX-1 is close to star forming region and 
%has an optical counterpart consistent with a B supergiant/WR star, as well as scaling mass estimates, 
%\LEt{ Connect this to the previous paragraph. Done!} 
We find that the compact objects,  Swift~J1644+57 and  Swift~J2058+05  are  likely to be  supermassive BHs with  
$M_{BH} \geq 7\times 10^7 M_{\odot}$ and $\geq 2\times 10^7 M_{\odot}$, correspondingly. 
%$M_{BH}> 7\times10^6 M_{\odot}$. 

\begin{acknowledgements}
{This research was done using the data supplied by the UK $Swift$ Science Data Centre at the University of Leicester.}
We appreciate Cristiano Guidorzi who raised a fair question on the origin of  the TDE sources.   
Particularly, we also acknowledge extensive discussion and critical points by Sergio Campana on  the  paper content,   Pascal Chardonnet for useful  discussions and comments and Partick Aurenche for editing of the paper.  Finally,  we recognize   the thorough analysis  of the paper by an anonymous referee.  
%editing the text  of the paper by Partick Aurenche.
%We appreciate editing the text  of the paper by Mike Nowak and Tod  Strohmayer. 
%We also acknowledge the deep analysis of 
%the previous version of 
%the  paper  by the referee and the editor.
\end{acknowledgements}

\newpage

%%%%%%%%%%%%%%%%%%%%%%%%%%%%%%%%%%%%%%%%%%%%%%%%%%%%%%%%%%%%%%%%%%%%%%%
%
% Table 1 - Suzaku log
%
%%%%%%%%%%%%%%%%%%%%%%%%%%%%%%%%%%%%%%%%%%%%%%%%%%%%%%%%%%%%%%%%%%%%%%%
\begin{deluxetable}{l l l l l c c}
%%%%%\rotate
\tablewidth{0in}
\tabletypesize{\scriptsize}
    \tablecaption{List of the {\it Suzaku} %(``Sz''-sets) 
observations of Swift~J1644+57 in the 0.3 -- 10 keV  range   used in our analysis.}
    \renewcommand{\arraystretch}{1.2}
 \label{tab:list_SAX_Suzaku}      % is used to refer this table in the text
\tablehead
{Number of set & Obs. ID& Start time (UT)  &  End time (UT) & MJD interval & Mean count rate \\
              &        &                  &                &              &  (cts/s)      }%\\
% \hline                                   % inserts single horizontal line
\startdata
Sz1 ................ & 906001010$^{(1)}$      & 2011 April 6 02:24:43 & 2011 April 7 00:32:17 & 55657.1 -- 55658.0 & 4.59$\pm$0.03\\
Sz2 ................ & 707018010$^{(1)}$      & 2012 May 17 17:30:38  & 2012 May  19 01:15:11 & 56064.7 -- 56066.1 & 0.032$\pm$0.001\\
% \hline                                             %inserts single line
      \enddata
% \end{tabular}
% \tablebib{
\\References: 
(1) Usui \& Kawai (2015)
%}
% \end{table*}
\end{deluxetable}

%%%%%%%%%%%%%%%%%%%%%%%%%%%%%%%%%%%%%%%%%%%%%%%%%%%%%%%%%%%%%%%%%%%%%%%
%
% Table 2 - RXTE
%
%%%%%%%%%%%%%%%%%%%%%%%%%%%%%%%%%%%%%%%%%%%%%%%%%%%%%%%%%%%%%%%%%%%%%%%

\begin{deluxetable}{l l l l l c c}%{l l l l l c c}
%%%%%\rotate
\tablewidth{0in}
\tabletypesize{\scriptsize}
    \tablecaption{List of the {\it RXTE} observations of Swift~J1644+57 and Swift~J2058+05.    }
    \renewcommand{\arraystretch}{1.2}
 \label{tab:list_RXTE}      % is used to refer this table in the text
\tablehead
{Source & Number  & Obs. ID& Start time (UT)  &  End time (UT) & MJD interval \\% & Mean count rate \\
       & of set  &                  &                &              }%\\%&  (cts/s)      \\
%  Obs. ID& Start time (UT)  && End time (UT) &MJD interval \\    % table heading
% \hline                                   % inserts single horizontal line
\startdata
Sw~J1644+57 & R1 ........... & 96424-01-01-00 & 2011 March 30  04:53:04  &  2011 March 30  05:26:56  &55650.20 -- 55650.27 \\%$^{1,2}$ \\% & 0.145$\pm$0.003 \\
            & R2 ........... & 96424-01-01-01 & 2011 March 31  02:51:12  &  2011 March 31  04:09:36  &55651.11 -- 55651.17 \\%$^{1,2}$ \\% & 0.145$\pm$0.003 \\
            & R3 ........... & 96424-01-01-02 & 2011 March 31  04:39:12  &  2011 March 31  04:57:04  &55651.19 -- 55651.21 \\%$^{1,2}$ \\% & 0.145$\pm$0.003 \\
 \hline                                             %inserts single line
Sw~J2058+05 & R4 ........... & 96431-01-01-00 &  2011 June 1  01:45:36  & 2011 June 1  02:12:00  & 55713.07 -- 55713.09 \\%
      \enddata
% \hline                                             %inserts single line
% \end{tabular}
% \tablebib{
%References: 
%(1) Ravasio et al. (2001); % Ravasio, M., Tagliaferri, G., Ghisellini, G. et al., 2001, A\&A, 383, 763
%(2) Padovani et al. (2001). 
%(3) Webb et al. 2010, 2014; %, 2005;
%(4) Yan et al. 2015.
%}
% \end{table*} 
\end{deluxetable}

%%%%%%%%%%%%%%%%%%%%%%%%%%%%%%%%%%%%%%%%%%%%%%%%%%%%%%%%%%%%%%%%%%%%%%%
%
% Table 3 - Swift log
%
%%%%%%%%%%%%%%%%%%%%%%%%%%%%%%%%%%%%%%%%%%%%%%%%%%%%%%%%%%%%%%%%%%%%%%%
\begin{deluxetable}{l l l l l l c}%{l l l l l c c}%{l l l l l c c}
%%%%%\rotate
\tablewidth{0in}
\tabletypesize{\scriptsize}
    \tablecaption{List of the $Swift$ observations of Swift~J1644+57 and Swift~J2058+05. }
    \renewcommand{\arraystretch}{1.2}
 \label{tab:list_Swift}      % is used to refer this table in the text
\tablehead
{Source & Number  & Obs. ID& Start time (UT)  &  End time (UT) & MJD interval \\% & Mean count rate \\
       & of set  &                  &                &              }%\\%&  (cts/s)      \\
%              &        &                  &                &              &  (cts/s)      \\
% \hline                                   % inserts single horizontal line
\startdata
Sw~J1644 & Sw1 ................ & 00031955(002-106, 107-177, 179-181,  & 2011 Mar 31     & 2011 Dec 6   & 55651.0 -- 55901.0 \\%&\\
         &                      & 183-211, 213-255$^{1,2,3,4,5,6}$           &                 &              &                    \\%&\\
         &Sw2 ................ & 00032200(001-210, 212-217, 219,      & 2011 Dec 7      & 2012 Aug 15  & 55902.3 -- 56154.1 \\%&\\
         &                     & 221-225, 227-232, 234-237$^{1,2,3}$  &                 &              &                    \\%&\\
         &Sw3 ................ & 00032526(001-003, 005-110, 103-116,  & 2012 Aug 16     & 2015 Apr 24  & 56155.9 -- 57136.2 \\%&\\
         &                     & 118-255$^{2,3}$                      &                 &              &                    \\%&\\
         &Sw4 ................ & 00033765(001-005, 007-060, 062-079)$^{}$            & 2015 May 1   & 2016 Oct 21 & 57143 -- 57682 \\%&\\
         &Sw5 ................ & 00450158(000-001, 004-007)$^{1,3,6}$     & 2011 Mar 28     & 2011 Mar 30  & 55648.6 -- 55650.6 \\%&\\
%  00092116(001-020)     & 2015 April 5       && 2015 Sept 22         & 57117.7 -- 57287.1 &\\
%  00359182000       & 2006 Aug. 29 11:38:56 && 2006 Aug. 29 21:24:57 & 53976.8 -- 53976.9 &\\
 \hline                                             %inserts single line
Sw~J2058 & Sw6 ................ & 00032004001$^{7,9}$             &  2011 May 27     & 2011 May 30   & 55708.9 -- 55709.4 \\%&\\
         & Sw7 ................ & 000320260(03-07, 09-16, 18-21)$^{7,8,9}$  & 2011 July 11    & 2011 Dec 7   & 55753.5 -- 55902.3 \\%&\\
      \enddata
% \hline                                             %inserts single line
% \end{tabular}
% \tablebib{
\\References: %Sw J1644 
(1) Saxton et al. (2012); 
(2) Komossa et al. (2015); 
(3) Mangano et al. (2016); %.%, 2014; %, 2005;
(4) Burrows et al. (2011);
(5) Levan et al. (2011);
(6) Zauderer et al. (2013); 
% Sw J2058:
(7) Pasham et al. (2015); 
(8) Komossa et al. (2015); 
(9) \cite{Cenko12}. %Mangano et al. (2016); %.%, 2014; %, 2005;
%}
% \end{table*}
\end{deluxetable}

%%%%%%%%%%%%%%%%%%%%%%%%%%%%%%%%%%%%%%%%%%%%%%%%%%%%%%%%%%%%%%%%%%%%%%%
%
% Table 4 - Suzaku spectral analysis
%
%%%%%%%%%%%%%%%%%%%%%%%%%%%%%%%%%%%%%%%%%%%%%%%%%%%%%%%%%%%%%%%%%%%%%%%

\begin{deluxetable}{llll}  %
%%%%%\rotate
\tablewidth{0in}
\tabletypesize{\scriptsize}
    \tablecaption{Best-fit parameters  of the  $Suzaku$ spectra of Swift J1644+57 in the 0.45$-$10~keV range using 
 the following four %two 
models$^\dagger$: phabs*zphabs*powerlw, phabs*zphabs*zbbody, phabs*zphabs(zbbody+zpowerlw) and phabs*zphabs*bmc.}
    \renewcommand{\arraystretch}{1.2}
    \label{tab:par_suz}
\tablehead
{     & Parameter & 906001010 & 707018010 \\%& 5088100200 & 5116500100 & 5116500110 \\%& Band-F & Band-G \\% & Band-C & Band-D \\
%     &           & Sz1       & Sz2       \\%& 5088100200 & 5116500100 & 5116500110 \\%& Band-F & Band-G \\% & Band-C & Band-D \\
%     & Parameter & S1 & S2 & S3 & S4 & S5 \\%& Band-F & Band-G \\% & Band-C & Band-D \\
% \hline                                             %inserts single line
% \hline                                             %inserts single line
Model &      &        &       }%\\% &       &      &    } \\% &     &   \\% & Band-C & Band-D \\
% \hline                                             %inserts single line
\startdata
phabs       & N$_{H, Gal}$    & 0.02$^f$    & 0.02$^f$    \\%& 6.8$\pm$0.09  & 7.6$\pm$0.2 & 2.9$\pm$0.6 \\% & 5.1$\pm$0.1  & 5.08$\pm$0.09 \\%& 0.03$\pm$0.01 & 0.05$\pm$0.01 & 0.03$\pm$0.01 \\
zphabs      & N$_{H,z=0.354}$ & 0.9$\pm$0.1 & 0.8$\pm$0.1 \\%& 6.8$\pm$0.09  & 7.6$\pm$0.2 & 2.9$\pm$0.6 \\% & 5.1$\pm$0.1  & 5.08$\pm$0.09 \\%& 0.03$\pm$0.01 & 0.05$\pm$0.01 & 0.03$\pm$0.01 \\
zpowerlw    & $\Gamma_{zpow}$ & 1.93$\pm$0.01 & 1.3$\pm$0.1 \\%& 1.71$\pm$0.04 & 1.94$\pm$0.09 & 2.61$\pm$0.04  \\% & 2.8$\pm$0.3 & 3.9$\pm$0.4 \\% & 2.0$\pm$0.2  & 1.4$\pm$0.2 \\
            & N$_{zpow}^{\dagger\dagger}$ & 33.2$\pm$0.5 & 0.11$\pm$0.03 \\%& 302$\pm$30 & 210$\pm$55 & 1686$\pm$110 \\% & 4.3$\pm$0.1 & 6.2$\pm$0.1 \\%& 0.31$\pm$0.05 & 0.04$\pm$0.01\\
      \hline
           & $\chi^2$ {\footnotesize (d.o.f.)} & 2.12 (1568)     & 1.25 (177)     \\%& 1.3 (79) & 0.72 (79)   & 1.05 (116) \\%& 3.2 (93)     & 2.1 (219)    \\% & 1.8 (19)     & 1.1 (19)  \\
      \hline
%Bbody      & T$_{BB}$ (keV)    & 65$\pm$2   & 70$\pm$3 & 85$\pm$3     & 94$\pm$4          & 64$\pm$5   & 72$\pm$4 & 83$\pm$6     & 95$\pm$7\\
phabs       & N$_{H, Gal}$    & 0.02$^f$    & 0.02$^f$    \\%& 6.8$\pm$0.09  & 7.6$\pm$0.2 & 2.9$\pm$0.6 \\% & 5.1$\pm$0.1  & 5.08$\pm$0.09 \\%& 0.03$\pm$0.01 & 0.05$\pm$0.01 & 0.03$\pm$0.01 \\
zphabs      & N$_{H,z=0.354}$ & 0.21$\pm$0.04 & 0.2$\pm$0.1 \\%& 6.7$\pm$0.5  & 6.8$\pm$0.3 & 2.9$\pm$0.08 \\% & 4.9$\pm$0.2  & 5.03$\pm$0.06 \\%& 0.03$\pm$0.01 & 0.05$\pm$0.01 & 0.03$\pm$0.01 \\
zbbody      & kT$_{zbb}$  (keV)   & 1.27$\pm$0.01   & 1.9$\pm$0.1  \\%& 1.19$\pm$0.01    & 1.00$\pm$0.01  & 0.67$\pm$0.01 \\%  & 85$\pm$7   & 120$\pm$10 \\%  & 85$\pm$3  & 72$\pm$5\\
%          & N$_{BB}^{\dagger\dagger}$ & 1.07$\pm$0.02  & 0.009$\pm$0.003   & 2.1$\pm$0.5   & 3.1$\pm$0.5 & 3.5$\pm$0.4  \\% & 4.1$\pm$0.5 & 5.0$\pm$0.3  \\%  & 1.8$\pm$0.6 & 1.5$\pm$0.4\\
           & N$_{zbb}^{\dagger\dagger}$ & 1.07$\pm$0.02  & 0.009$\pm$0.003 \\%& 15.4$\pm$0.3 & 7.5$\pm$0.3 & 32.9$\pm$0.6 \\%  & 4.3$\pm$0.1 & 6.2$\pm$0.1 \\%& 0.31$\pm$0.05 & 0.04$\pm$0.01\\
      \hline
           & $\chi^2$ {\footnotesize (d.o.f.)} &  5.07 (1568) & 1.54 (177) \\%& 2.31 (79)& 3.06 (79)& 13.44 (116) \\%& 1.19 (91) & 1.14 (217) \\
%           & $\chi^2$ {\footnotesize (d.o.f.)} & 1.14 (18)     & 1.28 (18)     & 1.94 (18)   & 1.93 (18)    & 3.03 (18) & 1.02 (19)     & 1.14 (19)   \\%  & 1.7 (19)     & 3.1 (19)\\
      \hline
%Bbody      & T$_{BB}$ (keV)    & 70$\pm$3   & 86$\pm$4   & 89$\pm$3  & 70$\pm$4         & 71$\pm$5   & 83$\pm$6   & 85$\pm$3  & 72$\pm$5\\
phabs       & N$_{H, Gal}$    & 0.02$^f$    & 0.02$^f$    \\%& 6.8$\pm$0.09  & 7.6$\pm$0.2 & 2.9$\pm$0.6 \\% & 5.1$\pm$0.1  & 5.08$\pm$0.09 \\%& 0.03$\pm$0.01 & 0.05$\pm$0.01 & 0.03$\pm$0.01 \\
zphabs      & N$_{H,z=0.354}$ & 1.24$\pm$0.03 & 1.9$\pm$0.4 \\%& 2.7$\pm$0.6  & 4.7$\pm$0.4 & 3.0$\pm$0.1 \\% & 5.1$\pm$0.1  & 5.08$\pm$0.09 \\%& 0.03$\pm$0.01 & 0.05$\pm$0.01 & 0.03$\pm$0.01 \\
zbbody      & kT$_{zbb}$ (keV)  & 0.29$\pm$0.02 & 0.24$\pm$0.03 \\% & 1.02$\pm$0.03  & 0.64$\pm$0.16  & 0.37$\pm$0.05     \\%    & 92$\pm$5   & 110$\pm$8 \\%  & 85$\pm$3  & 72$\pm$5\\
           & N$_{zbb}^{\dagger\dagger}$ & 0.16$\pm$0.09  & 0.03$\pm$0.01 \\%  & 3.8$\pm$0.3   & 1.4$\pm$0.1 & 8.7$\pm$1.2  \\% & 4.9$\pm$0.6 & 5.0$\pm$0.2  \\%  & 1.8$\pm$0.6 & 1.5$\pm$0.4\\
zpowerlw  & $\Gamma_{pow}$    & 1.92$\pm$0.02 & 1.3$\pm$0.1 \\%& 1.46$\pm$0.03  & 1.6$\pm$0.3 & 2.4$\pm$0.1 \\% & 2.6$\pm$0.1 & 2.9$\pm$0.3 \\%& 1.6$\pm$0.2 & 3.4$\pm$0.3\\
%Power-law  & $\Gamma_{pow}$    & 2.2$\pm$0.1 & 2.1$\pm$0.4 & 1.5$\pm$0.1 & 3.4$\pm$0.3 & 2.2$\pm$0.1 & 2.1$\pm$0.4 & 1.6$\pm$0.2 & 3.4$\pm$0.3\\
           & N$_{pow}^{\dagger\dagger}$ & 32.5$\pm$0.8 & 0.01$\pm$0.07 \\%& 150$\pm$10   & 108$\pm$80 & 1290$\pm$220 \\% & 0.67$\pm$0.02& 0.52$\pm$0.03 \\%& 0.12$\pm$0.07 & 1.3$\pm$0.4 \\
      \hline
           & $\chi^2$ {\footnotesize (d.o.f.)}& 1.03 (1566) & 1.17 (175) \\%& 1.78 (77) & 0.75 (77)& 0.97 (114) \\%& 1.19 (91) & 1.14 (217) \\% & 1.09 (17)& 1.2 (17)\\
      \hline
phabs       & N$_{H, Gal}$    & 0.02$^f$    & 0.02$^f$    \\%& 6.8$\pm$0.09  & 7.6$\pm$0.2 & 2.9$\pm$0.6 \\% & 5.1$\pm$0.1  & 5.08$\pm$0.09 \\%& 0.03$\pm$0.01 & 0.05$\pm$0.01 & 0.03$\pm$0.01 \\
zphabs      & N$_{H,z=0.354}$ & 1.56$\pm$0.01 & 1.5$\pm$0.2 \\%& 6.9$\pm$0.1  & 7.4$\pm$0.3 & 2.9$\pm$0.1 \\%& 4.9$\pm$0.1  & 5.02$\pm$0.04 \\%& 0.03$\pm$0.01 & 0.05$\pm$0.01 & 0.03$\pm$0.01 \\
bmc$^z$    & $\Gamma_{BMC}$    & 1.76$\pm$0.01 & 1.2$\pm$0.1 \\%& 2.21$\pm$0.09  \\% & 2.1$\pm$0.2    & 2.2$\pm$0.2  \\%  & 2.96$\pm$0.09    & 3.0$\pm$0.1 \\%& 2.5$\pm$0.3    & 2.1$\pm$0.2  & 1.7$\pm$0.1    & 1.4$\pm$0.1 \\    
           & kT$_{s}$   (eV)    & 189$\pm$3       & 270$\pm$7 \\%& 71$\pm$10       & 48$\pm$7       & 50$\pm$10    \\%& 105$\pm$9      & 130$\pm$10   \\%  & 56$\pm$10      & 43$\pm$8\\
           & logA$$            & -0.7$\pm$0.2  & -0.3$\pm$0.2 \\%& -0.72$\pm$0.09    & -0.86$\pm$0.04  & 0.35$\pm$0.07 \\% & -1.06$\pm$0.05   & -0.7$\pm$0.3 \\%& -4.3$\pm$0.4   & -3.9$\pm$0.5 \\
           & N$_{bmc}^{\dagger\dagger}$ & 0.47$\pm$0.07 & 0.0002$\pm$0.0001 \\%& 0.98$\pm$0.07  & 2.5$\pm$0.3  & 3.08$\pm$0.06 \\% & 4.03$\pm$0.05& 5.2$\pm$0.1 \\%& 1.4$\pm$0.1 & 0.6$\pm$0.1 \\
      \hline
           & $\chi^2$ {\footnotesize (d.o.f.)}& 0.98 (1566) & 0.97 (175) \\%& 1.17 (77)& 0.95 (77)& 1.08 (114) \\% & 0.79 (91) & 1.14 (217) \\% & 1.09 (17)& 1.2 (17)\\
%      \hline
      \enddata
% \hline                                             %inserts single line
% \end{tabular}
%\tablefoot{ 
$^\dagger$     Errors are given at the 90\% confidence level. %this parameter is fixed, 
\\$^{\dagger\dagger}$ The normalization parameters of blackbody and BMC components are in units of $L^{soft}_{36}/d^2_{10}$ erg s$^{-1}$ kpc$^{-2}$, 
where $L^{soft}_{36}$ is the soft photon luminosity in units of $10^{36}$ erg s$^{-1}$, $d_{10}$ is the distance to the 
source in units of 10 kpc, and power-law component is in units of %10$^{-6}$ %photons 
keV$^{-1}$ cm$^{-2}$ s$^{-1}$ at 1 keV. 
%$^{\dagger\dagger\dagger}$ spectral flux 
%in the 3-- 150 energy range 
%in units of $\times 10^{-15}$ erg/s/cm$^2$. 
%$N_H$ is the column density for the neutral absorber, {in units of $10^{20}$ cm$^{-2}$} (see details in the text).
% $5\times 10^{20}$ cm$^{-2}$ (see details in the text). %   in units of $10^{22}$ cm$^{-2}$. %($\times 10^{22}$ cm$^{-2}$).
$kT_{BB}$ and $kT_{s}$ are the temperatures of 
the blackbody and seed photon components (in keV and eV), respectively. 
$\Gamma_{pow}$ and $\Gamma_{BMC}$ are the indices of the { zpowerlw} 
and bmc, respectively. Hereafter superscript $^z$ indicates that the cosmological reddening of the spectrum was talen into ccount 
[$E\to E(1+z)$]. Redshift z was fixed to 0.354 (Levan et al. 2011). $N_{H, Gal}$ and $N_{H, z=0.354}$ are in units of $10^{22}$ cm$^{-2}$. 
%$^{\dagger\dagger\dagger}$ Flux is the unabsorbed flux of the continuum between 0.3 and 7 keV in units of 
%$10^{-15}$ erg s$^{-1}$ cm$^{-2}$. 
$^f$ indicates that a parameter was fixed.
%}
% \end{table}
% \end{table*}
\end{deluxetable}

%%%%%%%%%%%%%%%%%%%%%%%%%%%%%%%%%%%%%%%%%%%%%%%%%%%%%%%%%%%%%%%%%%%%%%%
%
% Table 5 - RXTE ANALYSIS
%
%%%%%%%%%%%%%%%%%%%%%%%%%%%%%%%%%%%%%%%%%%%%%%%%%%%%%%%%%%%%%%%%%%%%%%%
\begin{deluxetable}{lllllllllr}  %
%%%%%\rotate
\tablewidth{0in}
\tabletypesize{\scriptsize}
    \tablecaption{Best-fit parameters  of the $RXTE$ spectra of Swift~J1644+57 and Swift~J2058+05$^{a}$.}
    \renewcommand{\arraystretch}{1.2}
    \label{tab:par_rxte}
\tablehead
{
Source & Number of set &  $\alpha=\Gamma-1$ & $kT_s$ (eV) & $\log$(A) & $N^b$ & $E_{cut}$ (keV) & $kT_{e}$ (keV) & $\chi^2_{red}$ (d.o.f.) }%\\%& Flux$^d$ \\% & Band-C & Band-D \\%          &                &                    &  (eV) &              &              &        \\%&  \\%(0.4 -- 10 keV)\\% & Band-C & Band-D \\
\startdata
Sw~J1644 & R1$^c$ ................   & 0.76$\pm$0.01 & 90$^e$  & 0.5$\pm$0.3 & 1.1$\pm$0.1 & 90 $\pm$ 4 & ..... & 0.90 (74) \\ %&   38.50 \\ % 3.69    18.17  11.47  34.81 4.083   0.027 53450
&R2$^c$ ................   & 0.73$\pm$0.02 & 90$^e$  & -0.26$\pm$0.08 & 0.6$\pm$0.2 & 100 $\pm$ 7 & ..... & 0.95 (74) \\ %&   38.50 \\ % 3.69    18.17  11.47  34.81 4.083   0.027 53450
&R3$^c$ ................   & 0.78$\pm$0.04 & 90$^e$  & -0.3$\pm$0.1 & 0.5$\pm$0.1 & 110 $\pm$ 8 & ..... & 0.95 (74) \\ %&   38.50 \\ % 3.69    18.17  11.47  34.81 4.083   0.027 53450
&R1+Sz1$^d$ .......   & 0.79$\pm$0.05 & 370 $\pm$10   & -0.43$\pm$0.04 & 0.4$\pm$0.1 & ..... & 34 $\pm$ 2 &  1.03 (899) \\ %&   38.50 \\ % 3.69    18.17  11.47  34.81 4.083   0.027 53450
 \hline                                             %inserts single line
Sw~J2058 & R4$^c$ ................ & 0.6$\pm$0.1   & 160$^e$     &   0.3$\pm$0.1  & 3.4$\pm$0.5 & 16 $\pm$ 4 & ..... & 0.93 (125) \\ %
&R4+Sw6$^d$ .......      & 0.45$\pm$0.06 & 210 $\pm$30 & -0.43$\pm$0.04 & 3.8$\pm$0.3 & ..... & 4.0 $\pm$ 0.8 &  1.02 (1094) \\ %
% \hline                                             %inserts single line
% \hline                                             %inserts single line
      \enddata
% \end{tabular}
%\tablefoot{ 
\\$^a$  Errors are given at the 90\% confidence level. %this parameter is fixed, 
$^{b}$ The normalization parameters of the BMC and COMPTB components are in units of 
$L^{soft}_{36}/d^2_{10}$ erg s$^{-1}$ kpc$^{-2}$, 
where $L^{soft}_{36}$ is the soft photon luminosity in units of $10^{36}$ erg s$^{-1}$, $d_{10}$ is the distance to the 
source in units of 10 kpc. %, and Power-law component is in units of 10$^{-6}$ %photons 
%$^c$ 
$N_{H, Gal}$, the Galactic column density for the neutral absorber, was fixed to $2\times 10^{20}$ cm$^{-2}$ (for Sw~J1644+57) 
and $6.5\times 10^{20}$ cm$^{-2}$ (for Sw~J2058+05; Kalberla et al., 2005)  and 
$N_{H, z=0.354}$/$N_{H, z=1.1853}$, the column density for the neutral absorber in the local frame at redshift $z=0.354$/$z=1.1853$, was fixed to 
 $1.5\times 10^{22}$ cm$^{-2}$ (for Sw~J1644+57; Saxton et al. 2012) and $2.6\times 10^{21}$ cm$^{-2}$ (for Sw~J2058+05; Cenko et al. 2012), respectively; %, {in units of $10^{22}$ cm$^{-2}$} see details in the text).
$kT_{s}$ is the seed photon temperature, %(in eV), %$\Gamma_{pow}$ and $\Gamma_{bmc}$ are the indices of the {\it power law} 
$kT_e$ is the electron temperature, and $E_{cut}$ is the cutoff energy.
%$^{d}$ The unabsorbed flux of the continuum between 0.4 and 10 keV in units of $10^{-12}$ erg s$^{-1}$ cm$^{-2}$,  
$^c$ spectrum was fitted by the phabs*zphabs*bmc$^z$*zhighect model;
%$^d$ spectrum was fitted by phabs*zphabs*CompTB$^z$*zhighect model;
$^d$ spectrum was fitted by the phabs*zphabs*CompTB$^z$ model;  
$^e$ indicates that a parameter was fixed. % (see comments in the text). 
%}
%% \end{tabular}
%%\tablefoot{ 
%%$^a$     Errors are given at the 90\% confidence level. %this parameter is fixed, 
%%$^{b}$ The normalization parameters of BMC and COMPTB components are in units of 
%%$L^{soft}_{36}/d^2_{10}$ erg s$^{-1}$ kpc$^{-2}$, 
%%where $L^{soft}_{36}$ is the soft photon luminosity in units of $10^{36}$ erg s$^{-1}$, $d_{10}$ is the distance to the 
%%source in units of 10 kpc. %, and Power-law component is in units of 10$^{-6}$ %photons 
%$^c$ 
%%
%%$N_{H, Gal}$, the Galactic column density for the neutral absorber, was fixed to $2\times 10^{20}$ cm$^{-2}$  and 
%%$N_{H, z=0.354}$, the column density for the neutral absorber in the local frame at redshift $z=0.354$, was fixed to 
%%$1.5\times 10^{22}$ cm$^{-2}$ (Saxton et al. 2012); %, {in units of $10^{22}$ cm$^{-2}$} see details in the text).
%%$kT_{s}$ is the seed photon temperature, %(in eV), %$\Gamma_{pow}$ and $\Gamma_{bmc}$ are the indices of the {\it power law} 
%%$kT_e$ is the electron temperature, and $E_{cfe}$ is the cutoff energy.
%%%$^{d}$ The unabsorbed flux of the continuum between 0.4 and 10 keV in units of $10^{-12}$ erg s$^{-1}$ cm$^{-2}$,  
%%$^c$ spectrum was fitted by phabs*zphabs*bmc model;
%%$^d$ spectrum was fitted by phabs*zphabs*CompTB$^z$*zhighect model; 
%%$^e$ indicates that a parameter was fixed. % (see comments in the text).
%%}
% \end{table}
% \end{table*}
\end{deluxetable}
%%%%%%%%%%%%%%%%%%%%%%%%%%%%%%%%%%%%%%%%%%%%%%%%%%%%%%%%%%%%%%%%%%%%%%%
%
% Table 6 - SWIFT ANALYSIS
%
%%%%%%%%%%%%%%%%%%%%%%%%%%%%%%%%%%%%%%%%%%%%%%%%%%%%%%%%%%%%%%%%%%%%%%%
\begin{deluxetable}{llll}
%%%%%\rotate
\tablewidth{0in}
\tabletypesize{\scriptsize}
    \tablecaption{Best-fit parameters  of the  Swift spectra of Swift~J2058+05 in the 0.3$-$10~keV range using 
 the following four %two 
models$^\dagger$: phabs*zphabs*powerlw$, $phabs*zphabs*zbbody, phabs*zphabs(zbbody+zpowerlw) and phabs*zphabs*bmc.}
    \renewcommand{\arraystretch}{1.2}
    \label{tab:par_sw_20}
\tablehead
{
     & Parameter & 00032004001 & 00032026004 \\%
 \hline                                             %inserts single line
% \hline                                             %inserts single line
Model &      &        &       }%\\% &       &      &     \\% &     &   \\% & Band-C & Band-D \\
% \hline                                             %inserts single line
\startdata
phabs       & N$_{H, Gal}$    & 0.65$^f$    & 0.65$^f$    \\%
zphabs      & N$_{H,z=1.185}$ & 3.7$\pm$0.8 & 3.0$\pm$0.1 \\%&
zpowerlw    & $\Gamma_{zpow}$ & 1.69$\pm$0.08 & 1.3$\pm$0.1 \\%
            & N$_{zpow}^{\dagger\dagger}$ & 35.1$\pm$0.9 & 0.41$\pm$0.06 \\%
      \hline
           & $\chi^2$ {\footnotesize (d.o.f.)} & 3.2 (967)     & 1.27 (967)     \\%
      \hline
phabs       & N$_{H, Gal}$    & 0.65$^f$    & 0.65$^f$    \\%
zphabs      & N$_{H,z=1.185}$ & 0.2$\pm$0.3 & 0.6$\pm$0.1 \\%
zbbody      & kT$_{zbb}$  (keV)   & 3.89$\pm$0.08   & 1.6$\pm$0.1  \\%
            & N$_{zbb}^{\dagger\dagger}$ & 3.7$\pm$0.1  & 0.007$\pm$0.001 \\%
      \hline
           & $\chi^2$ {\footnotesize (d.o.f.)} &  5.07 (967) & 1.54 (967) \\%
      \hline
phabs       & N$_{H, Gal}$    & 0.65$^f$    & 0.65$^f$    \\%
zphabs      & N$_{H,z=1.185}$ & 1.9$\pm$0.2 & 2.3$\pm$0.5 \\%
zbbody      & kT$_{zbb}$ (keV)  & 5.7$\pm$0.6 & 1.2$\pm$0.3 \\% 
            & N$_{zbb}^{\dagger\dagger}$ & 13.6$\pm$0.2  & 0.3$\pm$0.1 \\%  
zpowerlw    & $\Gamma_{pow}$    & 1.92$\pm$0.02 & 1.3$\pm$0.1 \\%
            & N$_{pow}^{\dagger\dagger}$ & 210$\pm$10 & 15$\pm$9 \\%
      \hline
           & $\chi^2$ {\footnotesize (d.o.f.)}& 1.03 (969) & 1.17 (969) \\%
      \hline
phabs       & N$_{H, Gal}$     & 0.65$^f$      & 0.65$^f$     \\%
zphabs      & N$_{H,z=1.185}$  & 2.6$\pm$0.1   & 2.5$\pm$0.2  \\%
bmc$^z$    & $\Gamma_{BMC}$    & 1.78$\pm$0.03 & 1.5$\pm$0.2  \\%
           & kT$_{s}$   (eV)    & 300$\pm$10    & 270$\pm$7    \\%
           & log A$$            & 0.3$\pm$0.1   & 1.25$\pm$0.6 \\%
           & N$_{bmc}^{\dagger\dagger}$ & 3.1$\pm$0.1 & 0.09$\pm$0.01 \\%
      \hline
           & $\chi^2$ {\footnotesize (d.o.f.)}& 1.04 (969) & 0.89 (969) \\%
%      \hline
      \enddata
% \hline                                             %inserts single line
% \end{tabular}
%\tablefoot{ 
\\$^\dagger$     Errors are given at the 90\% confidence level. %this parameter is fixed, 
$^{\dagger\dagger}$ The normalization parameters of the blackbody and BMC components are in units of $L^{soft}_{35}/d^2_{10}$ erg s$^{-1}$ kpc$^{-2}$, 
where $L^{soft}_{35}$ is the soft photon luminosity in units of $10^{35}$ erg s$^{-1}$, $d_{10}$ is the distance to the 
source in units of 10 kpc, and power-law component is in units of %10$^{-6}$ %photons 
keV$^{-1}$ cm$^{-2}$ s$^{-1}$ at 1 keV. 
%$^{\dagger\dagger\dagger}$ spectral flux 
%in the 3-- 150 energy range 
%in units of $\times 10^{-15}$ erg/s/cm$^2$. 
%$N_H$ is the column density for the neutral absorber, {in units of $10^{20}$ cm$^{-2}$} (see details in the text).
% $5\times 10^{20}$ cm$^{-2}$ (see details in the text). %   in units of $10^{22}$ cm$^{-2}$. %($\times 10^{22}$ cm$^{-2}$).
k$T_{BB}$ and k$T_{s}$ are the temperatures of 
the blackbody and seed photon components (in keV and eV), respectively. 
$\Gamma_{pow}$ and $\Gamma_{BMC}$ are the indices of the  zpowerlw 
and bmc, respectively. Hereafter superscript $^z$ indicates that the cosmological reddening of the spectrum was taken into account 
[$E\to E(1+z)$]. Redshift z was fixed to 1.185~\citep{Cenko12}. $N_{H, Gal}$ and $N_{H, z=1.185}$ are in units of $10^{21}$ cm$^{-2}$. 
%$^{\dagger\dagger\dagger}$ Flux is the unabsorbed flux of the continuum between 0.3 and 7 keV in units of 
%$10^{-15}$ erg s$^{-1}$ cm$^{-2}$. 
$^f$ indicates that a parameter was fixed.

%}
% \end{table}
% \end{table*}
\end{deluxetable}

%%%%%%%%%%%%%%%%%%%%%%%%%%%%%%%%%%%%%%%%%%%%%%%%%%%%%%%%%%%%%%%%%%%%%%%
%
% Table 7 - Parametrization
%
%%%%%%%%%%%%%%%%%%%%%%%%%%%%%%%%%%%%%%%%%%%%%%%%%%%%%%%%%%%%%%%%%%%%%%%
\begin{deluxetable}{lcccccc}
%%%%%\rotate
\tablewidth{0in}
\tabletypesize{\scriptsize}
    \tablecaption{Parameterizations for  the target sources, Swift~J1644+57 and Swift~J2058+05, and reference sources.}
    \renewcommand{\arraystretch}{1.2}
 \label{tab:parametrization_scal}
\tablehead
{
  Reference source  &       $\cal A$ &     $\cal B$     &  $\cal D$  &    $x_{tr}$      & $\beta$  &  }%\\
%      \hline
\startdata
GRO~J1655--40  & 1.98$\pm$0.02 &  0.44$\pm$0.02  &  1.0 & 0.06$\pm$0.02   &   1.88$\pm$0.25  \\
GX~339--4 RISE 2004     & 2.24$\pm$0.01 &  0.51$\pm$0.02  &  1.0 & 0.039$\pm$0.002   &   3.5  \\
4U~1543--37  DECAY 2002 & 2.15$\pm$0.06 &  0.63$\pm$0.07  &  1.0 & 0.049$\pm$0.001   &   0.6$\pm$0.1  \\
Cyg~X--1      &  2.11$\pm$0.06 &  0.59$\pm$0.08  &  1.0 &  0.076$\pm$0.003   &     0.8$\pm$0.1  \\
%
%XTE~J1550-564 RISE 1998 & 2.84$\pm$0.08 &  1.8$\pm$0.3    &  1.0 & 0.132$\pm$0.004   &   0.61$\pm$0.02  \\
%H~1743-322    RISE 2003 & 2.97$\pm$0.07 &  1.27$\pm$0.08  &  1.0 & 0.053$\pm$0.001   &   0.62$\pm$0.04  \\
%4U~1630-472   & 2.88$\pm$0.06 &  1.29$\pm$0.07  &  1.0 & 0.045$\pm$0.002   &   0.64$\pm$0.03  \\
%M101 ULX-1   & 2.88$\pm$0.06 &  1.29$\pm$0.07   & 1.0  &   [4.2$\pm$0.2]$\times 10^{-4}$ &   0.61$\pm$0.03  \\%& $constant$ \\
 \hline\hline                        % inserts double horizontal lines
%                    &               &           &      &                             &         & \\  
%  Target source     &       A       &     B     &   D  &   $x_{tr} [\times 10^{-5}]$ & $\beta$ \\%& $N_H$ hypothesis \\
  Target source     &      $\cal A$     &    $\cal B$    &  $\cal  D$  &   $x_{tr} [\times 10^{-6}]$ & $\beta$ \\%& $N_H$ hypothesis \\
      \hline
Swift~J1644+57 & 1.84$\pm$0.09 & 0.46$\pm$0.08   & 1.0  &   6.59$\pm$0.09 & 1.78$\pm$0.08  \\%& $variable$ \\
Swift~J2058+05 & 1.84$\pm$0.09 & 0.46$\pm$0.08   & 1.0  &   6.59$\pm$0.09 & 1.78$\pm$0.08  \\%& $variable$ \\
%BL Lacertae & 2.17$\pm$0.09 & 0.62$\pm$0.07   & 1.0  &   9.58$\pm$0.06 & 0.51$\pm$0.06  \\%& $variable$ \\
%M101 ULX-1   & 2.88$\pm$0.06 &  1.29$\pm$0.07   & 1.0  &   [4.2$\pm$0.2]$\times 10^{-4}$ &   0.61$\pm$0.03  \\%& $constant$ \\
%ESO 243-49 HLX-1 & 3.00$\pm$0.04 & 1.27$\pm$0.05   & 1.0  &   4.25$\pm$0.03 & 0.62$\pm$0.05  \\%& $variable$ \\
%ESO 243-49 HLX-1 & 3.00$\pm$0.04 & 1.27$\pm$0.04   & 1.0  &   [4.25$\pm$0.03]$\times 10^{-6}$ & 0.62$\pm$0.04  \\%& $variable$ \\
% \hline                                             %inserts single line
      \enddata
% \end{tabular}
% \end{table*}
\end{deluxetable}

%%%%%%%%%%%%%%%%%%%%%%%%%%%%%%%%%%%%%%%%%%%%%%%%%%%%%%%%%%%%%%%%%%%%%%%
%
% Table 8 - BH mass determination
%
%%%%%%%%%%%%%%%%%%%%%%%%%%%%%%%%%%%%%%%%%%%%%%%%%%%%%%%%%%%%%%%%%%%%%%%
\begin{deluxetable}{llllccc}
%%%%%\rotate
\tablewidth{0in}
\tabletypesize{\scriptsize}
    \tablecaption{Esimates of a BH mass and a distance  value for the target sources, Swift~J1644+57 and Swift~J2058+05, and those for reference   sources. }
    \renewcommand{\arraystretch}{1.2}
 \label{tab:par_scal}
\tablehead
{
      Source   & M$^{\dagger}_{dyn}$ (M$_{\odot})$ & i$_{orb}^{\dagger}$ (deg) & d$^{\dagger\dagger}$ (kpc)  & $M_{fund.plane}$ (M$_{\odot}$) & $M_{min.timescale}$ (M$_{\odot}$) & M$_{scal}$ (M$_{\odot}$) }%\\%& $N_H$ hypothesis \\
%      \hline
\startdata
GX~339-4$^{(1, 2)}$    &   $> 6c^{(1)}$        &     ...       &  7.5$\pm$1.6$^{(2)}$      &...& ...&   5.7$\pm$0.8$^{\dagger\dagger\dagger}$ \\%&  1, 2, 3 \\
4U~1543-47$^{(3, 4)}$   &   9.4$\pm$1.0 &  20.7$\pm$1.5 &  7.5$\pm$1.0           &...& ...&   9.4$\pm$1.4$^{\dagger\dagger\dagger}$ \\%&  1, 2, 3 \\
Cyg~X--1$^{(5)}$     &   6.8-13.3  &  35$\pm$5   &  2.5$\pm$0.3     &...& ...&    7.9$\pm$1.0$^{\dagger\dagger\dagger}$ \\%&  1, 2, 3 \\
%4U~1543--47   &   9.4$\pm$1.0$^{(3, 4)}$ &  20.7$\pm$1.5$^{(5)}$ &  7.5$\pm$1.0$^{(3)}$, 9.1$\pm$1.1$^{(5)}$    &...&   9.4$\pm$1.4$^{\dagger\dagger\dagger}$ \\%&  1, 2, 3 \\
GRO~J1655--40   &   6.3$\pm$0.3$^{(7, 8)}$ &  70$\pm$1$^{(7, 8)}$ &  3.2$\pm$0.2$^{(9)}$    &... &...&   ... \\%&  1, 2, 3 \\
Swift~J1644+57$^{(10, 11, 12, 13, 14)}$  & ... &     ...      & $\sim$1.5$\times 10^6$ & $\sim$3$\times 10^6$& $\sim 10^7 - 10^9$ & $\ge 7.8\times10^{6}$ \\%& $constant$  \\% & 5, 6 and this work \\ %this work \\
Swift~J2058+05$^{(14, 15, 16)}$  & ... &     ...      & $\sim$ 3.7$\times 10^6$ & ... &$\sim$ 5$\times 10^7$ & $\ge 2 \times 10^{7}$ \\%& $constant$  \\%
%BL Lacertae$^{(6, 7, 8,9, 10, 11, 12, 13)}$  & ... &     ...      & $\sim$300$\times 10^3$ & $\sim$1.7$\times 10^8$& $\ge 3\times10^{7}$ \\%& $constant$  \\% & 5, 6 and this work \\ %this work \\
      \enddata
% \hline                                             %inserts single line
% \end{tabular}
% \tablebib{  
\\References:
(1) Munoz-Darias et al. (2008); 
(2) Hynes et al. (2004);
(3) Orosz  (2003); 
(4) Park et al. (2004); 
(5) Herrero et al. 1995; 
(6) Ninkov et al. 1987; 
%(5) Orosz et al. (1998);
%(6) Urry et al. (2000);
%(7) Ryle (2008);
%(8) Liang \& Liu (2003);
%(9) Oke \& Gumm (1974).
%(1) Orosz 2003; (2) Park et al. 2004; (Park, S. Q., et al. 2004, ApJ, 610, 378)
%(1) Orosz et al. 2002; 
%(2) S$\grave a$nchez-Fern$\grave a$ndez et al. 1999; 
%(3) Sobczak et al. 1999;  
%(4) Petri 2008; 
(7) Green et al. 2001; (8) Hjellming \& Rupen 1995; (9) Jonker \& Nelemans 2004. %, MNRAS, 354, 355; 
(10) Miller \& Gultekin (2011) %(Fund plane)
(11) Bloom et al. (2011); 
(12) Burrows et al. (2011); 
(13) Levan et al. (2011);
(14) Komossa (2015); 
%(8) Komossa et al. (2015); 
(15) Cenko et al. (2012);
(16) Pasham et al. (2015).
%(5) Mu$\grave n$oz-Darias et al. 2008; 
%(6) Hynes et al. 2004;
%(7) Orosz 2003; 
%(8) Park et al. 2004;
%(5) Urry et al., 2000;
%(6) Ryle, 2008.
%(4) McClintock et al. 2007;  
%(5) STS14;
%(6) Shappee \& Stanek 2011;
%(7) Mukai et al. 2005; 
%(8) Kelson et al. 1996;
%(9) TS15;
%}
%\tablefoot{ 
\\$^{\dagger}$ Dynamically determined BH mass and system inclination angle, $^{\dagger\dagger}$ Source distance found in the literature. 
For Swift~J2058+0516 the distance is estimated using the redshift z with taking into account the cosmological 
effects for $z > 1$ (see Wright 2006\footnote{http://www.astro.ucla.edu/~wright/CosmoCalc.html} ).%, 
$^{\dagger\dagger\dagger}$ Scaling value found by ST09. 
%} 
%\tablefoot{ 
%\\$^a$ Dynamically determined BH mass and system inclination angle, $^b$ Source distance found in literature,  
%$^c$ Scaling value found by ST09. 
%} 
% \end{table*}
\end{deluxetable}

\newpage
%%%%%%%%%%%%%%%%%%%%%%%%%%%%%%%%%%%%%%%%%%%%%%%%%%%%%%%%%%%%%%%%%%%%%%%%%%%%%%%%%%%%%%%%%%%%%%
% 
% Figure 1 LIGHT CURVE
%
%%%%%%%%%%%%%%%%%%%%%%%%%%%%%%%%%%%%%%%%%%%%%%%%%%%%%%%%%%%%%%%%%%%%%%%%%%%%%%%%%%%%%%%%%%%%%%

\begin{figure*}
%\begin{figure}
 \centering
\includegraphics[width=17cm]{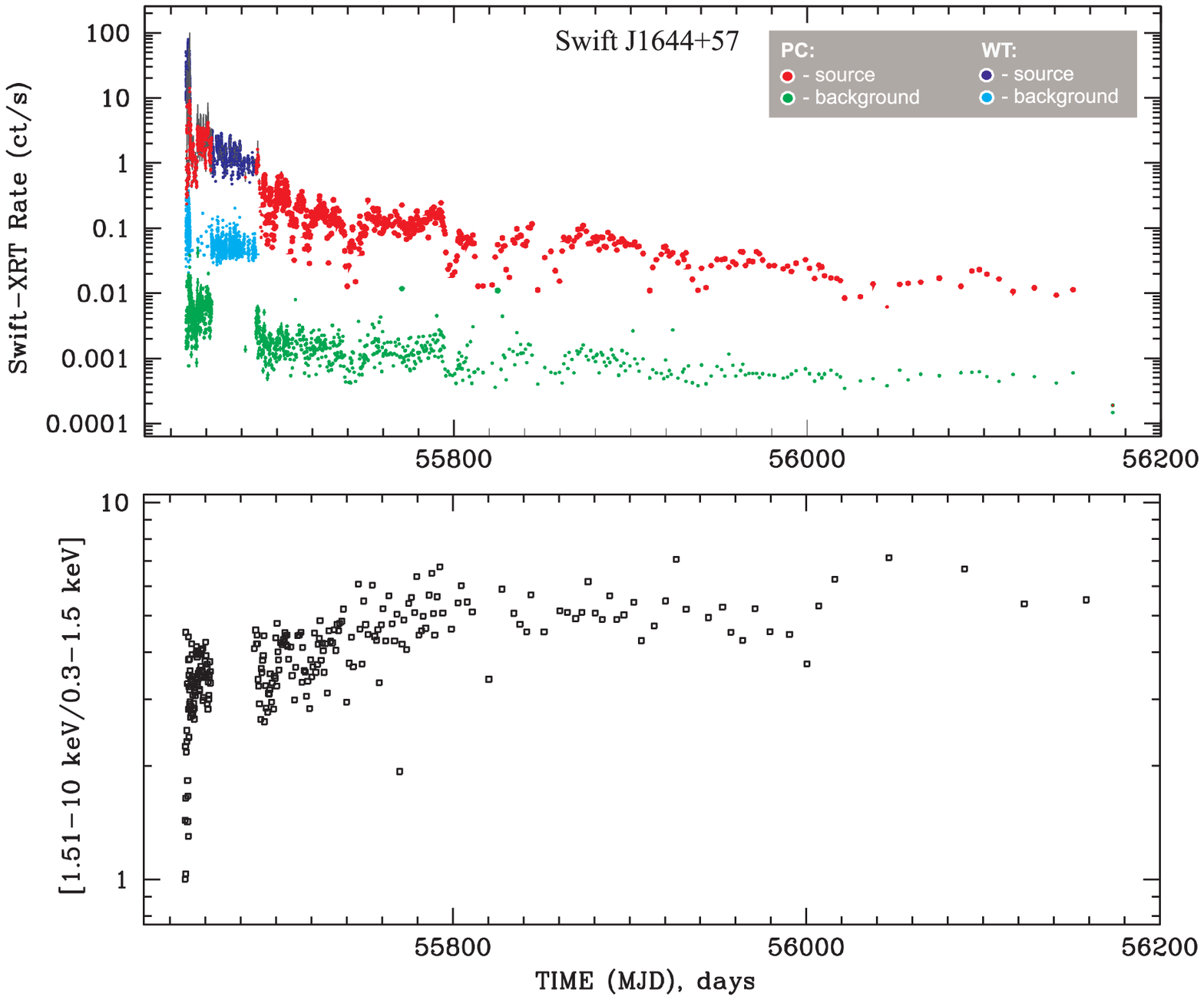}
\caption{
$Swift$/XRT light curve of Swift~J1644+57 in the 0.3 $-$ 10 keV  range during 2011 -- 2012 (top panel). Here, red, %\LEt{ or? The slash is ambiguous and should be replaced. LT: Agree!} 
blue points mark the source signal (with 2-$\sigma$ detection level) for PC/WT mode.  Green and cyan points  indicate the background level for PC/WT mode, respectively. %blue arrows  show the MJD of Suzaku.
In the bottom panel we show the hardness ratio curve (1.51 -- 10 keV/0.3 -- 1.5 keV)  using PC mode data of Swift~J1644+57. 
}
\label{lc}
 \end{figure*}

\newpage
%%%%%%%%%%%%%%%%%%%%%%%%%%%%%%%%%%%%%%%%%%%%%%%%%%%%%%%%%%%%%%%%%%%%%%%%%%%%%%%%%%%%%%%%%%%%%%
% 
% Figure 2 HID
%
%%%%%%%%%%%%%%%%%%%%%%%%%%%%%%%%%%%%%%%%%%%%%%%%%%%%%%%%%%%%%%%%%%%%%%%%%%%%%%%%%%%%%%%%%%%%%%
\begin{figure*}
%\begin{figure}
 \centering
\includegraphics[width=18cm]{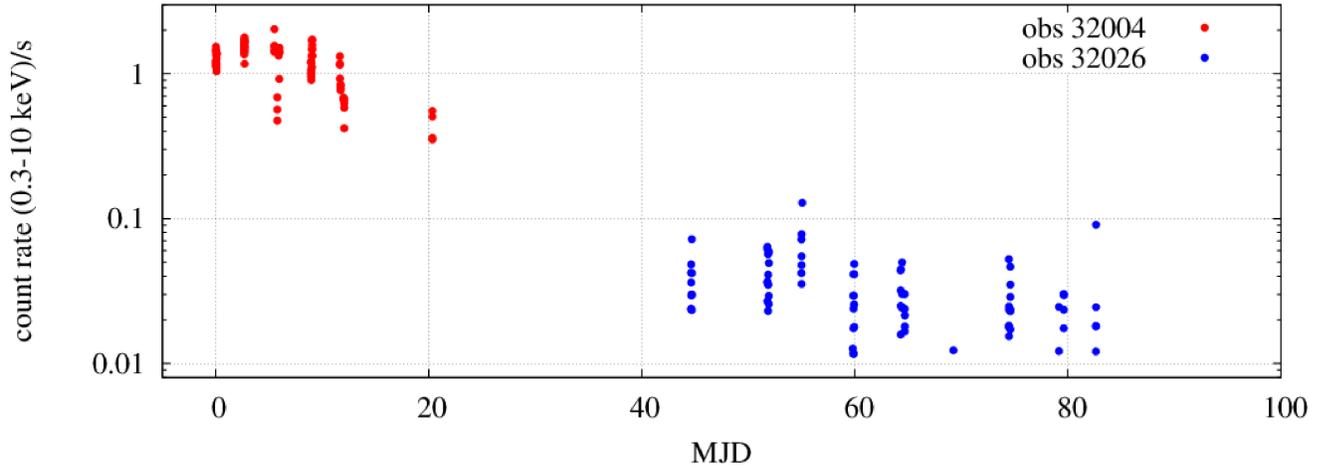}
\caption{Light curve of Swift~J2058+05 in the 0.3 -- 10 keV energy range during the outburst decay (2011, May, 27 -- July, 11) deteted by $Swift$/XRT. 
The red points mark the brighter phase (id=32004) of outburst, while the blue points indicate the fainter phase (id=32026). 
{In the horizontal axis we show the time scale in the MJD-55708.9 units} % .}
}
\label{lc_20}
 \end{figure*}
% \end{figure}

\newpage
%%%%%%%%%%%%%%%%%%%%%%%%%%%%%%%%%%%%%%%%%%%%%%%%%%%%%%%%%%%%%%%%%%%%%%%%%%%%%%%%%%%%%%%%%%%%%%
% 
% Figure 3 
%
%%%%%%%%%%%%%%%%%%%%%%%%%%%%%%%%%%%%%%%%%%%%%%%%%%%%%%%%%%%%%%%%%%%%%%%%%%%%%%%%%%%%%%%%%%%%%%
 \begin{figure*}
%  \begin{figure}
 \centering
    \includegraphics[width=14cm]{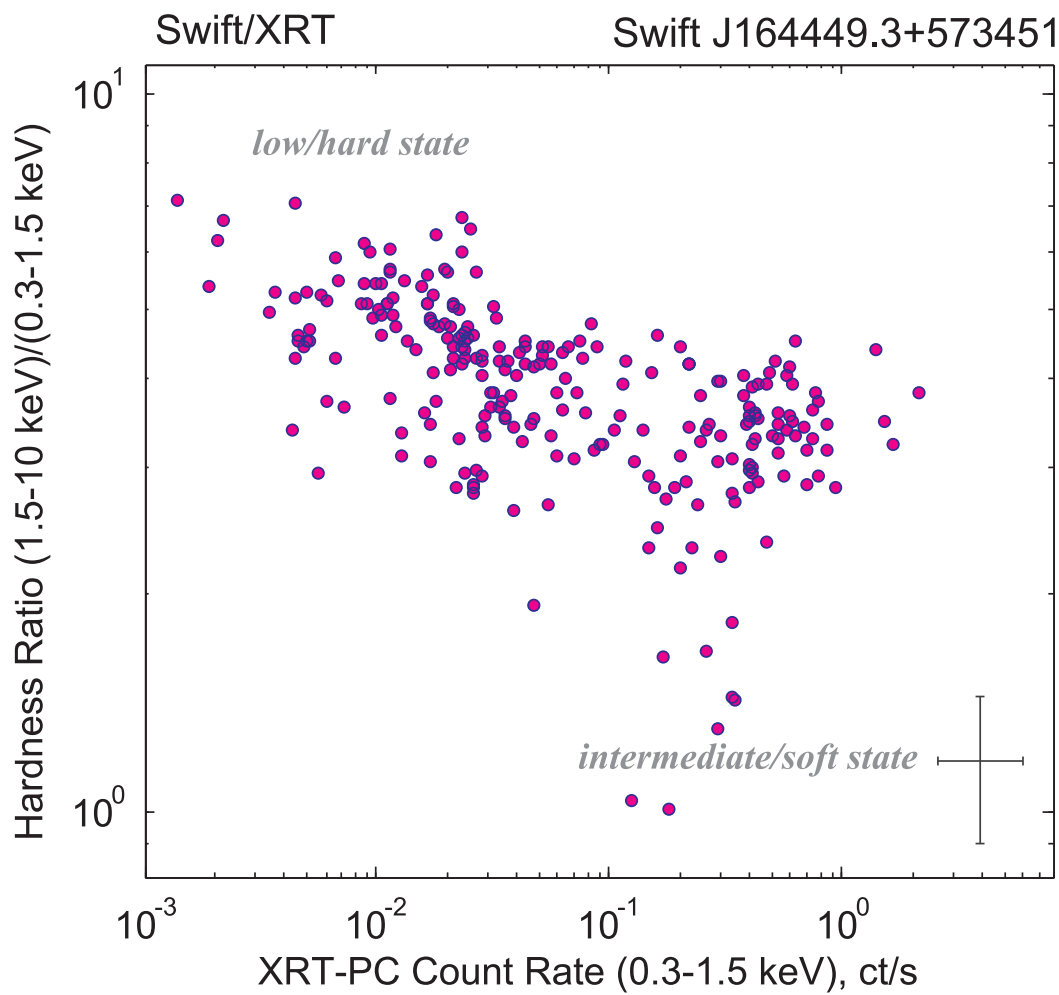}
      \caption{ 
Hardness-intensity diagram (HID) for Swift~J1644+57 using the {\it Swift} observations (2011 -- 2012)   during spectral evolution from the {high} state to the {low} state. 
 In the vertical axis, the hardness ratio (HR)  is a ratio 
of the source counts in the  two energy bands: 
the {hard} (1.5 -- 10 keV) and  soft (0.3 -- 1.5 keV). 
HR  decreases with a source  brightness in the 0.3 -- 10 keV  range (horizontal axis). 
{ For clarity, we plot only one point with error bars (in the bottom right corner) to demonstrate 
typical uncertainties  for the count rate and HR.} % and .
}
   \label{HID}
% \end{figure*}
 \end{figure*}

\newpage
%%%%%%%%%%%%%%%%%%%%%%%%%%%%%%%%%%%%%%%%%%%%%%%%%%%%%%%%%%%%%%%%%%%%%%%%%%%%%%%%%%%%%%%%%%%%%%
% 
% Figure 4 
%
%%%%%%%%%%%%%%%%%%%%%%%%%%%%%%%%%%%%%%%%%%%%%%%%%%%%%%%%%%%%%%%%%%%%%%%%%%%%%%%%%%%%%%%%%%%%%%
\begin{figure*}
%\begin{figure}
 \centering
\includegraphics[width=16cm]{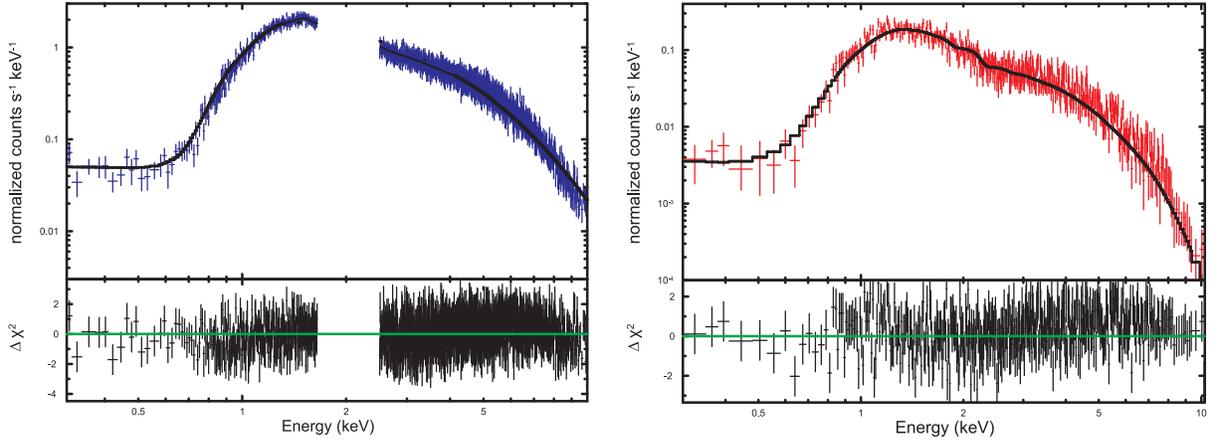}
\caption{
$Suzaku$/XIS spectra (blue) and $Swift$/XRT (red) of Swift~J1644+57 at the decay phase (both on 2011 April 6, MJD=55657) 
in normalized counts %/$EF(E)$ units  
fitted using the {phabs*zphabs*bmc$^z$} model   
with $\alpha=0.63\pm 0.01$, %$kT_e=25\pm 2$ keV, 
$\log(A)=-0.72\pm 0.2$, $kT_s=240\pm 5$ eV  for the $Suzaku$ spectrum, $Sz1$ ($\chi^2_{red}=0.98$ for 1566 d.o.f.) and with $\alpha=0.77\pm 0.04$, %$kT_e=25\pm 2$ keV, 
$\log(A)=-1.62\pm 0.09$, $kT_s=200\pm 20$ eV ($\chi^2_{red}=1.04$ for 659 d.o.f.) for the $Swift$ spectrum. 
%Here, %data are denoted by red points; 
The spectral models are %presented with components is 
shown by %red and 
$black$ hystograms. %, and dashed purple lines 
%for the $Swift$ and $Suzaku$ spectra, respectively.  
  }
\label{spectrum_Sz_sw}
 \end{figure*}

\newpage
%%%%%%%%%%%%%%%%%%%%%%%%%%%%%%%%%%%%%%%%%%%%%%%%%%%%%%%%%%%%%%%%%%%%%%%%%%%%%%%%%%%%%%%%%%%%%%
% 
% Figure 5 
%
%%%%%%%%%%%%%%%%%%%%%%%%%%%%%%%%%%%%%%%%%%%%%%%%%%%%%%%%%%%%%%%%%%%%%%%%%%%%%%%%%%%%%%%%%%%%%%
\begin{figure*}
%\begin{figure}
 \centering
\includegraphics[width=7cm]{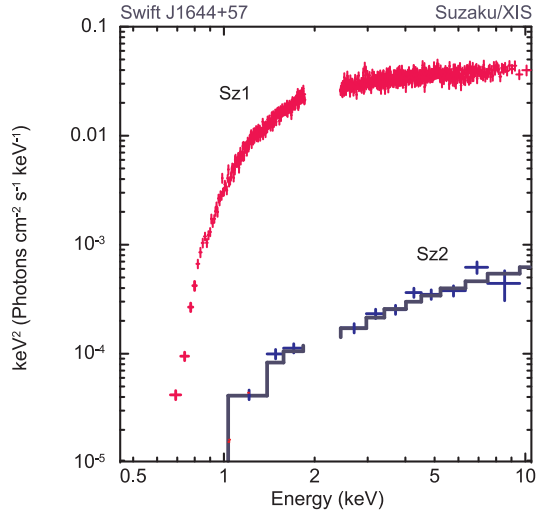}
\caption{
Two $E F_E$ spectral diagrams during the high soft
%\LEt{ replace the slash.}
 (purple) and low hard
%\LEt{ replace the slash.} 
(blue) spectral states of Swift~J1644+57. Data taken from 
$Suzaku$ observations, 906001010 (Sz1, the high soft
%\LEt{ replace the slash.}) 
and 707018010 (Sz2, the low hard).
%\LEt{ replace the slash.}).
}
\label{spectrum_Suz_70_90}
 \end{figure*}

\newpage
%%%%%%%%%%%%%%%%%%%%%%%%%%%%%%%%%%%%%%%%%%%%%%%%%%%%%%%%%%%%%%%%%%%%%%%%%%%%%%%%%%%%%%%%%%%%%%
% 
% Figure 6 RXTE spectrum E_curtoff~70 keV 
%
%%%%%%%%%%%%%%%%%%%%%%%%%%%%%%%%%%%%%%%%%%%%%%%%%%%%%%%%%%%%%%%%%%%%%%%%%%%%%%%%%%%%%%%%%%%%%%
\begin{figure*}
%\begin{figure}
 \centering
\includegraphics[width=14cm]{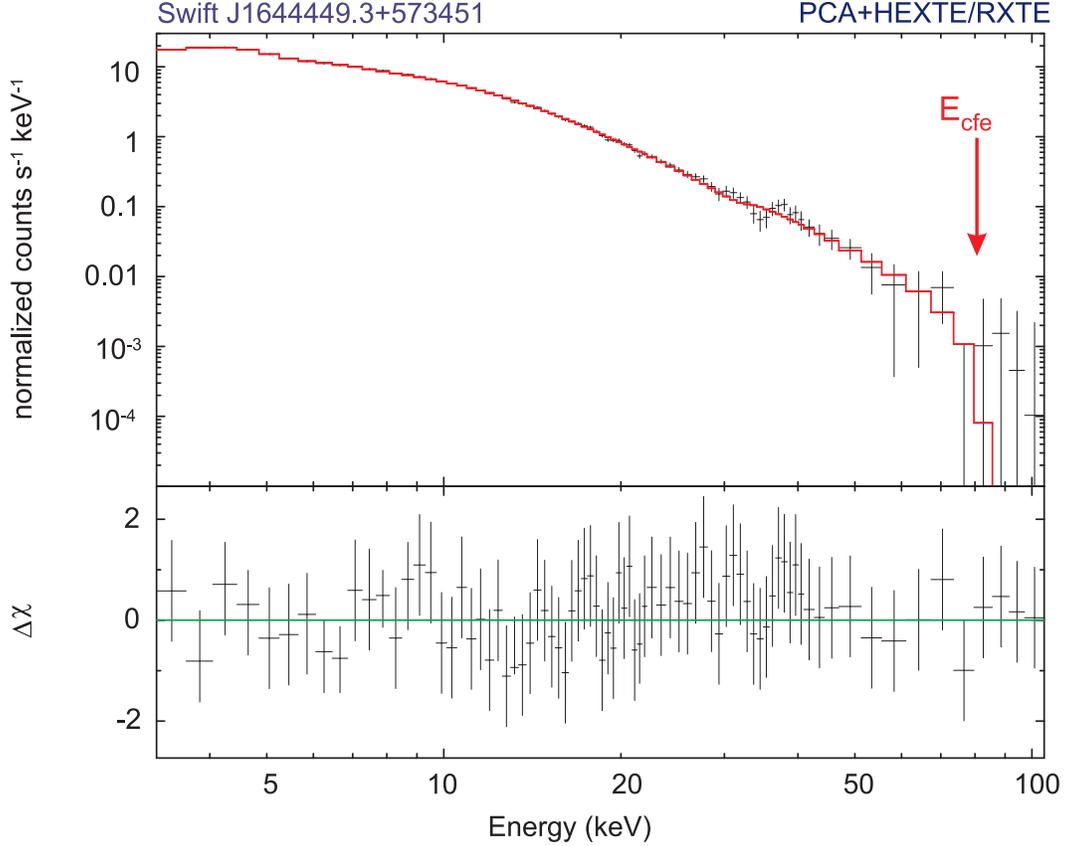}
\caption{
$RXTE$ (3 -- 100) keV spectrum of Swift J1644+57. The observation taken on 2011 March 30 (MJD=55650.0, ID=96242-01-01-00) 
was fitted using {\it CompTB} model with $\alpha=0.86\pm 0.03$, $kT_e=46\pm 3$ keV, $\log(A)=-0.5\pm 0.2$, 
$kT_s=0.3$ keV (fixed) 
{
with the intrinsic column density of $N_{H, z=0.354}=1.5\times 10^{22}$ cm$^{-2}$ (at redshift $z=0.354$) and 
the Galactic absorption $N_{H, Gal}=2\times 10^{20}$ cm$^{-2}$ 
}
 ($\chi^2_{red}=0.97$ for 120 d.o.f.).  The high energy cut-off is clearly seen at $E_{cut}=80$ keV.
%
%{\it Bulk motion Comptonization} model (with $\alpha=0.86\pm 0.01$) modified with high energy cut-off with 
%$E_{cut}=68\pm9$ keV.
}
\label{RXTE_spectrum_CompTB_E_cut_70keV}
\end{figure*}
% \end{figure}

\newpage
%%%%%%%%%%%%%%%%%%%%%%%%%%%%%%%%%%%%%%%%%%%%%%%%%%%%%%%%%%%%%%%%%%%%%%%%%%%%%%%%%%%%%%%%%%%%%%
% 
% Figure 7 
%
%%%%%%%%%%%%%%%%%%%%%%%%%%%%%%%%%%%%%%%%%%%%%%%%%%%%%%%%%%%%%%%%%%%%%%%%%%%%%%%%%%%%%%%%%%%%%%
\begin{figure*}
%\begin{figure}
 \centering
\includegraphics[width=14cm]{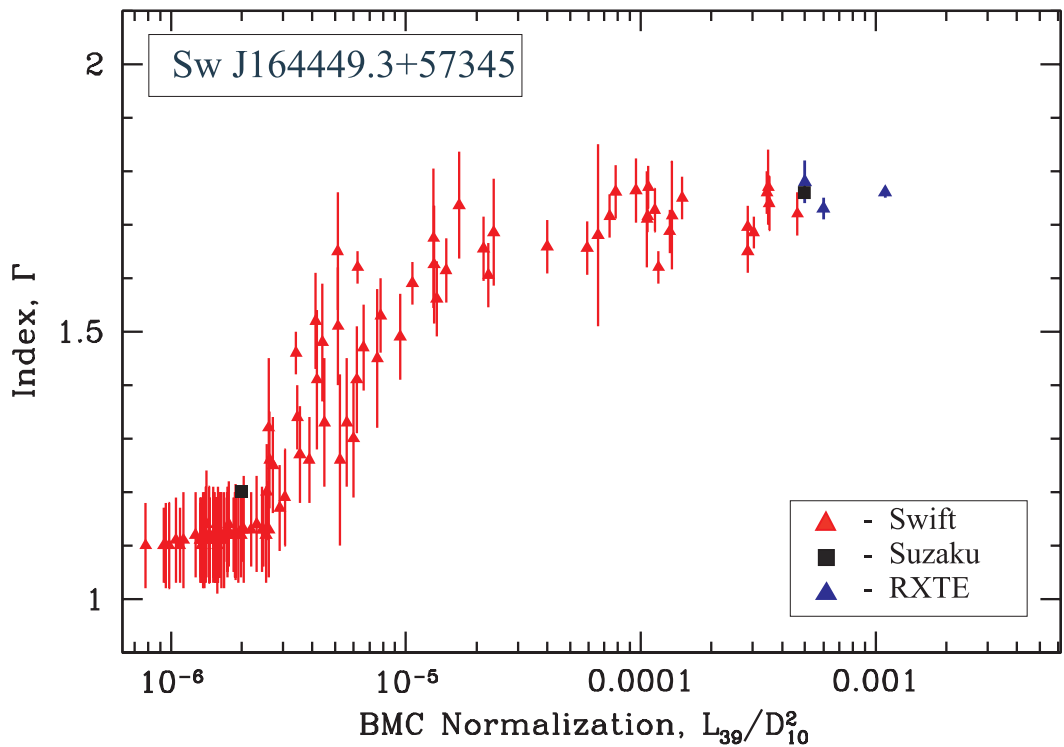}
\caption{
Photon index, $\Gamma$ plotted versus BMC normalization (which is proportional to $\dot M$)
for the TDE source, Swift J1644+57  using $Swift$ (red triangles), $Suzaku$ (black squares) and {\it RXTE} (blue 
triangles) data (see Tables~4-\ref{tab:par_rxte}).
}
%\label{gam vs norm}
\label{saturation}
 \end{figure*}

\newpage
%%%%%%%%%%%%%%%%%%%%%%%%%%%%%%%%%%%%%%%%%%%%%%%%%%%%%%%%%%%%%%%%%%%%%%%%%%%%%%%%%%%%%%%%%%%%%%
% 
% Figure 8 RADIO x-RAY CORRELATION
%
%%%%%%%%%%%%%%%%%%%%%%%%%%%%%%%%%%%%%%%%%%%%%%%%%%%%%%%%%%%%%%%%%%%%%%%%%%%%%%%%%%%%%%%%%%%%%%

\begin{figure*}
%\begin{figure}
\begin{center}
%   \resizebox{\hsize}{!}{\includegraphics{f7_tde.eps}}
 \includegraphics[width=16cm]{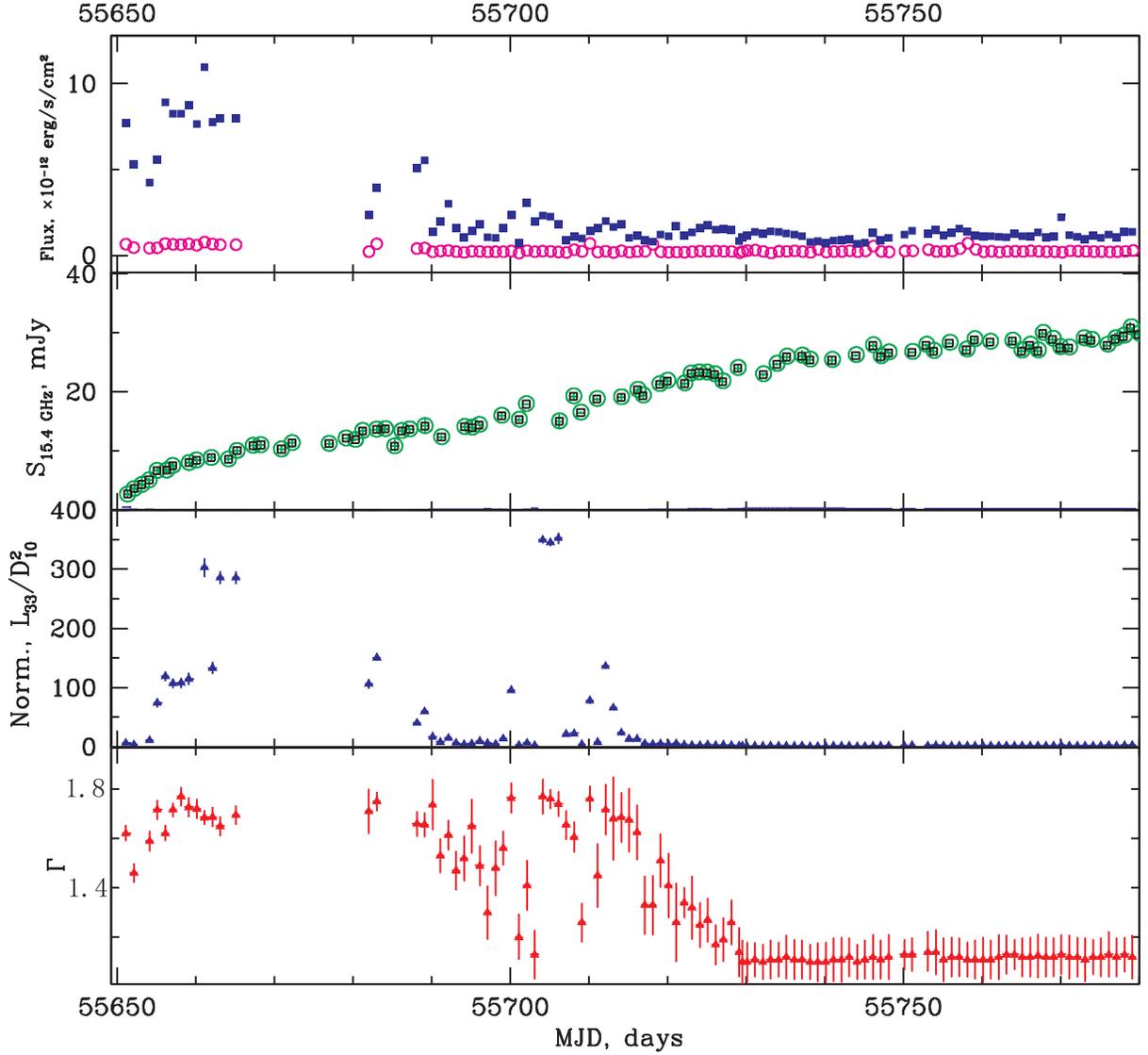} 
\end{center}
  
\caption{
{For Swift J1644+57, from top to bottom:}
evolutions of the model flux in the soft band  0.3 -- 1.5 keV, blue squares, % $black$ points, 
%$Swift$/XRT] 
and the hard band  1.5 -- 10 keV, pink circles ($Swift$/XRT bands); the flux density  
$S_{15.4 GHz}$ (green points) at 15.4 GHz (AMI Large Array, Berber et al. 2012), 
the BMC normalization and the photon index, $\Gamma$ during the 2011 flare decay set ({S1, S2}).
}
\label{radio}
\end{figure*}
\
%%%%%%%%%%%%%%%%%%%%%%%%%%%%%%%%%%%%%%%%%%%%%%%%%%%%%%%%%%%%%%%%%%%%%%%%%%%%%%%%%%%%%%%%%%%%%%
% 
% Figure 9 EF_E
%
%%%%%%%%%%%%%%%%%%%%%%%%%%%%%%%%%%%%%%%%%%%%%%%%%%%%%%%%%%%%%%%%%%%%%%%%%%%%%%%%%%%%%%%%%%%%%%

\begin{figure*}
%\begin{figure}
 \centering
\includegraphics[width=14cm]{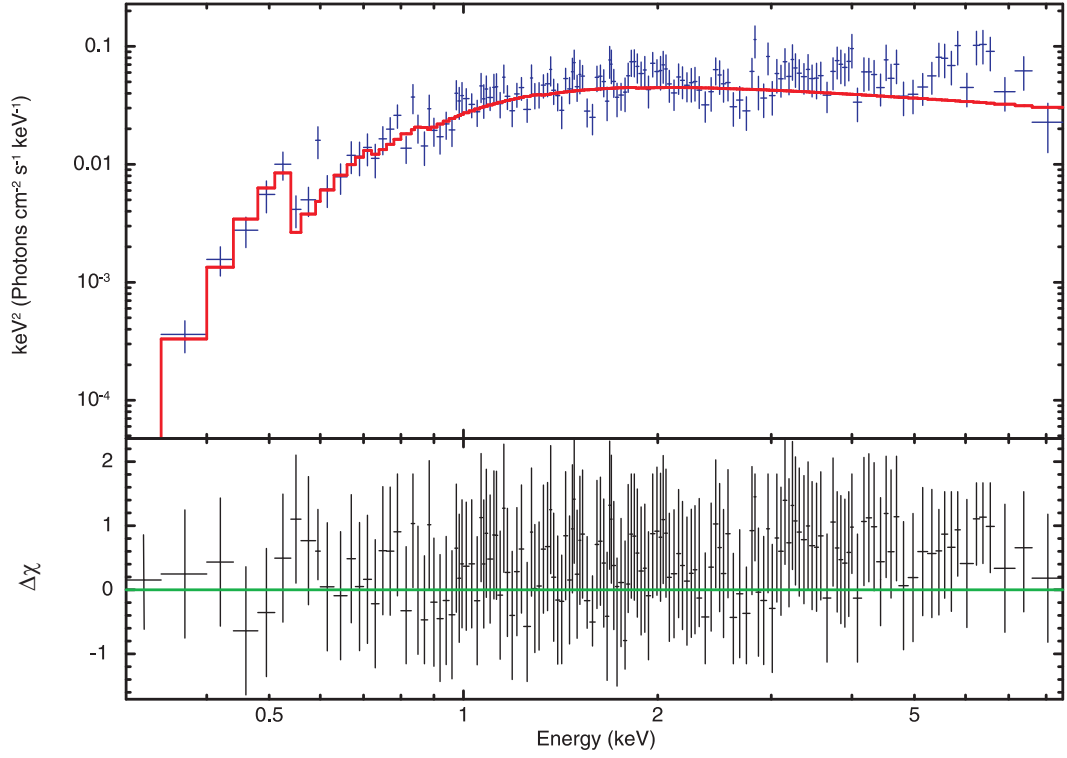}
\caption{Best-fit $Swift$/XRT spectrum in $EF_E$ units during  of the intermediate-low hard state  transition  (top panel) 
 with $\Delta\chi$ (bottom panel) for Swift~J2058+05.
}
\label{sw_spectrum}
 \end{figure*}
\newpage

%%%%%%%%%%%%%%%%%%%%%%%%%%%%%%%%%%%%%%%%%%%%%%%%%%%%%%%%%%%%%%%%%%%%%%%%%%%%%%%%%%%%%%%%%%%%%%
% 
% Figure 10  Three spectra
%
%%%%%%%%%%%%%%%%%%%%%%%%%%%%%%%%%%%%%%%%%%%%%%%%%%%%%%%%%%%%%%%%%%%%%%%%%%%%%%%%%%%%%%%%%%%%%%
\begin{figure*}
 \centering
    \includegraphics[width=12cm]{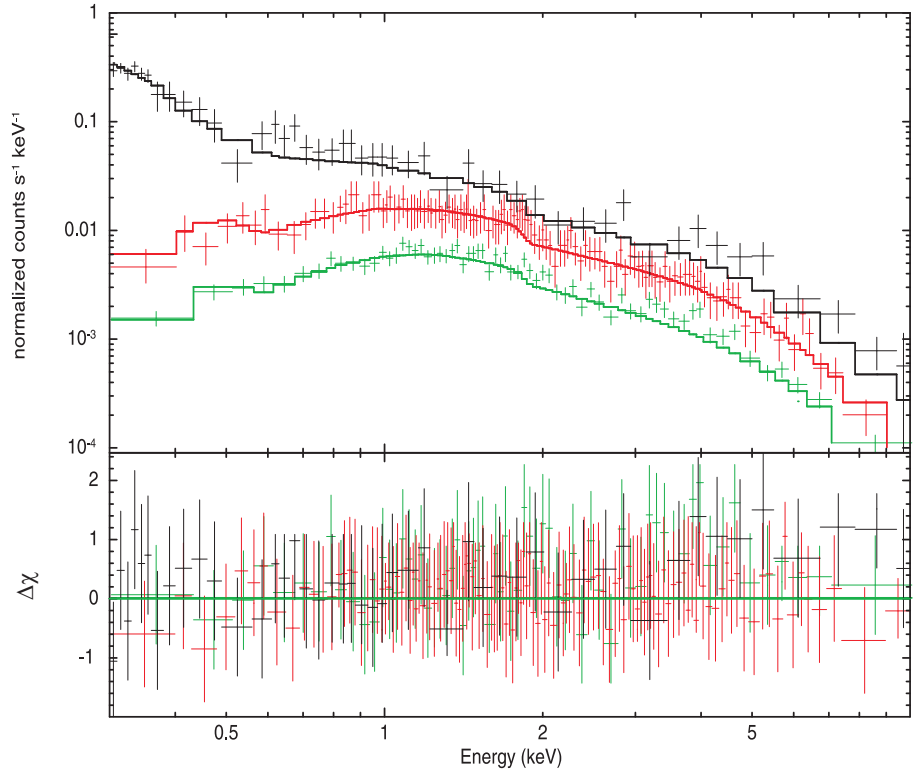}
      \caption{Three representative $Swift$/XRT spectra in normalized counts units (top panel) 
with $\Delta\chi$ (bottom panel) for the LHS ($green$), IS ($red$) and HSS ($black$) spectral states 
of Swift~J2058+05. 
%Hardness-intensity diagram (HID) for Swift~J2058+05 using the {\it Swift} observations (2011 -- 2012)   during spectral evolution from the {high} state to the {low} state. 
% In the vertical axis, the hardness ratio (HR)  is a ratio of the source counts in the  two energy bands: 
%the {hard} (1.5 -- 10 keV) and  soft (0.3 -- 1.5 keV). 
%HR  decreases with a source  brightness in the 0.3 -- 10 keV  range (horizontal axis). 
%{ For clarity, we plot only one point with error bars (in the bottom right corner) to demonstrate 
%typical uncertainties  for the count rate and HR.} % and .
}
   \label{three_sw_sp}
 \end{figure*}

\newpage

\begin{figure}
 %\centering
    \includegraphics[width=18cm]{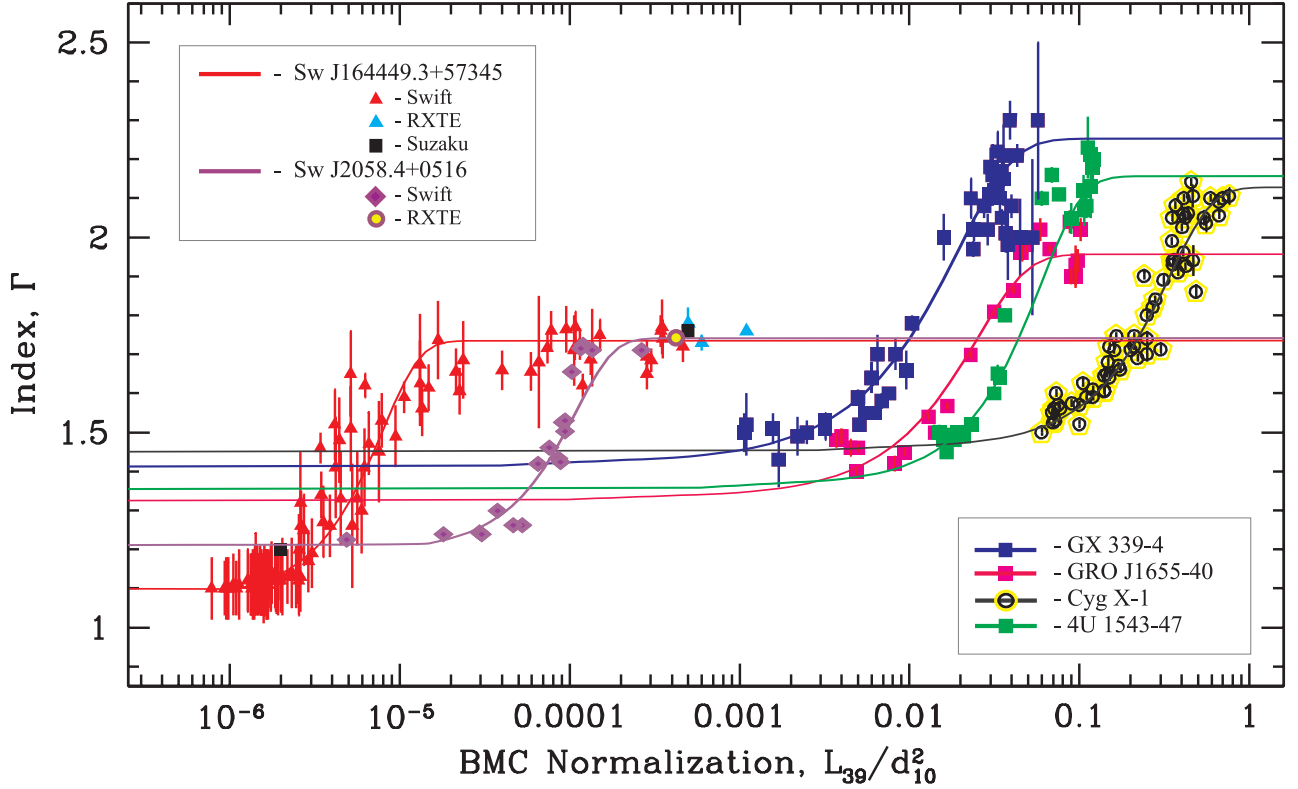}
      \caption{Scaling of the correlation of the photon index, $\Gamma$ versus the BMC normalization rate (which is proportional 
to mass accretion rate, $\dot M$) for Swift~J1644+57 (with red/black/bright blue points from $Swift/Suzaku/RXTE$ data) and 
Swift~J2058+05 (with violet diamonds/circle from $Swift/RXTE$ data),  and 
that for GRO~J1655--40 (with pink squares, ST09), GX~339--4 (with blue squares, ST09), Cyg~X--1 
(with black circles, ST09) and 4U~1543--47 (with green squares, ST09)  as a function of BMC normalization. 
}
\label{scaling}
\end{figure}

\end{document}